\documentclass[amsmath, amssymb, superscriptaddress]{revtex4-2}
\usepackage{graphicx}
\usepackage[english]{babel}
\usepackage[utf8]{inputenc}
\usepackage[T1]{fontenc}
\usepackage[
pdftitle={Eigenmodes of synthetic antiferromagnetic skyrmions},
pdfauthor={Kauser Zulfiqar, Martin Lang, Samuel J. R. Holt, Swapneel Amit Pathak,
  Florian Bruckner, and Hans Fangohr},
colorlinks=True,
citecolor=blue,
]{hyperref}

\usepackage{orcidlink}

\usepackage[colorinlistoftodos, color=green!40, prependcaption]{todonotes}

\begin{document}
\title{Eigenmodes of synthetic antiferromagnetic skyrmions}

\date{\today}

\author{Kauser Zulfiqar\,\orcidlink{0009-0004-5639-4114}}
\email[]{kauser.zulfiqar@mpsd.mpg.de}
\affiliation{Max Planck Institute for the Structure and Dynamics of Matter, Hamburg, Germany.}
\affiliation{Center for Free-Electron Laser Science, Hamburg, Germany.}
\affiliation{Department of Physics, University of Hamburg,
  Hamburg, Germany.}

\author{Martin Lang\,\orcidlink{0000-0001-7104-7867}}
\email[]{martin.lang@mpsd.mpg.de}
\affiliation{Max Planck Institute for the Structure and Dynamics of Matter, Hamburg, Germany.}
\affiliation{Center for Free-Electron Laser Science, Hamburg, Germany.}

\author{Samuel J. R. Holt\,\orcidlink{0000-0003-3323-8958}}
\affiliation{Max Planck Institute for the Structure and Dynamics of Matter, Hamburg, Germany.}
\affiliation{Center for Free-Electron Laser Science, Hamburg, Germany.}

\author{Swapneel Amit Pathak\,\orcidlink{0000-0003-3840-955X}}
\affiliation{Max Planck Institute for the Structure and Dynamics of Matter, Hamburg, Germany.}
\affiliation{Center for Free-Electron Laser Science, Hamburg, Germany.}

\author{Florian Bruckner\,\orcidlink{0000-0001-7778-6855}}
\affiliation{University of Vienna, Vienna, Austria}

\author{Hans Fangohr\,\orcidlink{0000-0001-5494-7193}}
\affiliation{Max Planck Institute for the Structure and Dynamics of Matter, Hamburg, Germany.}
\affiliation{Center for Free-Electron Laser Science, Hamburg, Germany.}
\affiliation{University of Southampton, Southampton, UK}

\begin{abstract}
We investigate the excitation modes of confined synthetic-antiferromagnetic
(SAF) skyrmions using micromagnetic eigenvalue and ringdown simulations.
Starting from a single skyrmion in a ferromagnetic layer, where the
lowest-frequency modes are a gyrotropic and a breathing mode, we study how
antiferromagnetic interlayer coupling modifies the dynamics in SAF bilayers. We
consider several geometries: single SAF skyrmions in square and rectangular
confinement, unequal layer thicknesses, and strips containing multiple
skyrmions.

The antiferromagnetic coupling strongly modifies the low-frequency dynamics. The
square geometry exhibits two nearly degenerate gyrotropic modes, where in each
both layers have the same rotation sense. In rectangular geometries, we instead
find nearly linear SAF skyrmion translation emerging from opposite gyration
sense in the two layers. These translational modes become the characteristic
low-frequency excitations of SAF skyrmion chains.

For skyrmion chains, we identify collective translational and breathing modes
with standing-wave-like spatial profiles. Beyond ferromagnetic-like breathing
modes, the SAF geometry supports breathing oscillations in which the two layers
oscillate out of phase. We further demonstrate signal propagation along extended
SAF skyrmion chains with propagation velocities comparable to ferromagnetic
skyrmion chains.

These results provide a systematic description of the collective dynamics of SAF
skyrmions arising from the interplay of geometric confinement, intralayer, and
interlayer coupling.
\end{abstract}


\maketitle

\section{Introduction}\label{sec:Introduction}
Magnetic skyrmions~\cite{skyrme1962a, rossler2006a, nagaosa2013d, finocchio2016, kong2019a,
  zhang2020a, zhang2023, song2020, yuan2022c}
are topologically non-trivial nanoscale spin textures with
potential for future spintronic and magnonic
applications~\cite{wolf2001, juarez-martinez2012, everschor-sitte2018, liu2020, hirohata2020, chen2023,  mishra2023, sud2025}.
These include skyrmion racetracks for non-volatile data storage~\cite{fert2013b, parkin2015a,
woo2016a}, skyrmion crystals for signal processing \cite{chumak2017, chen2021}, and magnonic logic
gates~\cite{luo2018,yan2021}.

Skyrmions, or a collection thereof, are a promising candidate for magnonic
applications because they host a rich variety of dynamic modes. These include
gyrotropic modes~\cite{buttner2015, guslienko2017} in which regions of the
skyrmion configuration orbit around an equilibrium position, and the breathing
modes~\cite{woo2017, garanin2020}, which can be characterised as a periodic
expansion and contraction of the configuration around the skyrmion center.
Skyrmion gyration and breathing modes, and their collective excitations have
been studied in detail~\cite{kim2017, kim2018a, garst2017collective}.

However, ferromagnetic skyrmions face several limitations that hinder their use
in practical devices. The non-local stray fields restrict their minimum size to
several tens of nanometres at room temperature~\cite{buttner2018}, and an
external magnetic field is commonly required to maintain their
stability~\cite{buttner2018, bernand-mantel2018, moreau-luchaire2016, woo2016a}
(though zero-field stability has been demonstrated~\cite{boulle2016, beg2017a}).

Another interesting class of materials that are known to host the skyrmion configuration are
synthetic antiferromagnets (SAFs)~\cite{zhao2023}. SAFs are multi-layer magnetic systems where
adjacent ferromagnetic layers are coupled antiferromagnetically due to Ruderman-Kittel-Kasuya-Yosida (RKKY)
interaction~\cite{duine2018}. The SAF system has two main advantages from the device application
point of view. Firstly, the absence of net magnetisation significantly reduces stray fields.
This reduction allows closer packing of magnetic components in the device,
and it also reduces the skyrmion size~\cite{buttner2015}.
Secondly, and relevant for systems that contain moving skyrmions, the
zero net topological charge of the SAF skyrmion cancels the skyrmion Hall effect and
simulations predict a straight trajectory along the current with
enhanced velocities compared to their ferromagnetic counterpart~\cite{zhang2016b}.
Recent experiments have also shown that skyrmions can be
stabilised at room temperature~\cite{legrand2020} in SAFs.

The dynamic modes of a single SAF skyrmion, in a disk-like geometry, have been
studied for both gyrotropic~\cite{xing2018} and breathing
modes~\cite{lonsky2020a}. Breathing modes of a vertical stack of skyrmions with
antiferromagnetic interlayer coupling have been investigated~\cite{barker2023}.

In this work, we extend these studies to incorporate multiple skyrmions in SAF
materials. A 1D ferromagnetic skyrmion chain has been studied
systematically~\cite{kim2017, kim2018a}. We instead consider a 1D SAF skyrmion
chain in a thin-film geometry with two antiferromagnetically coupled
ferromagnetic layers~\cite{lonsky2020a}. This allows us to simultaneously
investigate the intra-layer (skyrmion-skyrmion) and the inter-layer
(antiferromagnetic) coupling of skyrmions.

\section{Results} \label{sec:results}

Throughout this work we primarily focus on eigenmodes of (SAF) skyrmions. We use
the term \emph{SAF skyrmion} to refer to a stacked pair of antiferromagnetically
coupled skyrmions in the two layers of a synthetic antiferromagnet (SAF). The
equilibrium configurations are obtained with micromagnetic simulations and
eigenvalues and -vectors by solving the linearised LLG without damping. For a
few cases we also simulate the full dynamics of the system with an applied
external excitation, from which we can obtain a power spectral density (ringdown
method), for more details refer to the Methods section.

\begin{figure}
  \includegraphics[width=\linewidth]{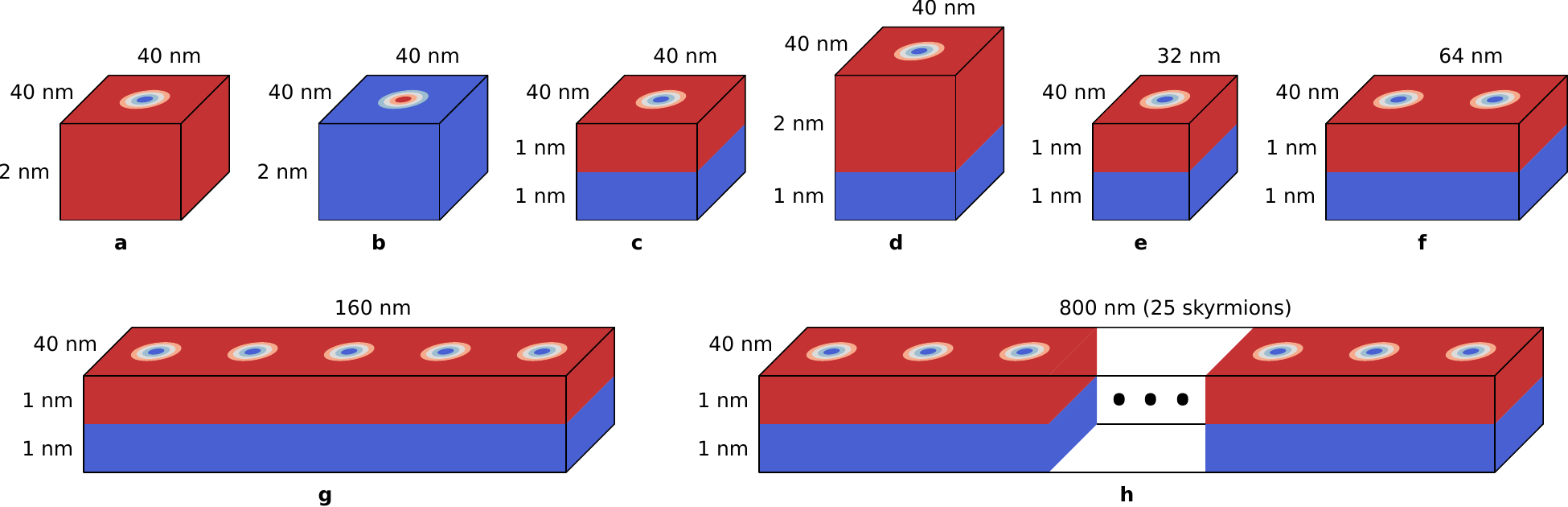}
  \caption{
    Schematic of geometries and skyrmion distribution studied in this work.
    Colour indicates the $m_\mathrm{z}$ component of the
    magnetisation (red $m_\mathrm{z}=+1$, blue $m_\mathrm{z}=-1$, grey
    $m_\mathrm{z}=0$). The skyrmions are schematically sketched on the top surface,
    they extend through the whole thickness. The skyrmion core polarisation is
    always opposite to the background magnetisation direction, i.e. flips at the
    layer interface for those systems where the background magnetisation is flipped.
    \label{fig:geometries}
  }
\end{figure}

Figure~\ref{fig:geometries} shows an overview of the different geometries
investigated in this work. Before we investigate one and more skyrmions in a SAF configuration, we will
first focus on a single skyrmion in a chiral ferromagnet (FM) with negative core
polarisation (Fig.~\ref{fig:geometries}a) and positive core polarisation
(Fig.~\ref{fig:geometries}b) in a 2\,nm film of square geometry
(Section~\ref{sec:one-skyrmion-fm}).

For the SAF studies (starting in Section~\ref{sec:one-skyrmion-saf}), we split
the 2\,nm thickness into 2 layers of 1\,nm each that are antiferromagnetically
coupled. We start with a single skyrmion in a square geometry
(Fig.~\ref{fig:geometries}c). We then extend the study for a single skyrmion to
uneven layer thickness (Fig.~\ref{fig:geometries}d) and rectangular geometry
(Fig.~\ref{fig:geometries}e). Finally, we extend the rectangular geometry along
the x direction such that the strip can host two (Fig.~\ref{fig:geometries}f),
five (Fig.~\ref{fig:geometries}g) or 25 (Fig.~\ref{fig:geometries}h) skyrmions.

The power spectral density depends (among other factors) on the spatial profile
of the excitation. For a single (SAF) skyrmion (Fig.~\ref{fig:geometries}a -- e)
we excite the whole sample. For more than one skyrmion we only excite the
left-most SAF skyrmion.

\subsection{Skyrmion in FM square geometry}\label{sec:one-skyrmion-fm}

\begin{figure}
  \sffamily\textbf a

	\includegraphics[width=\columnwidth]{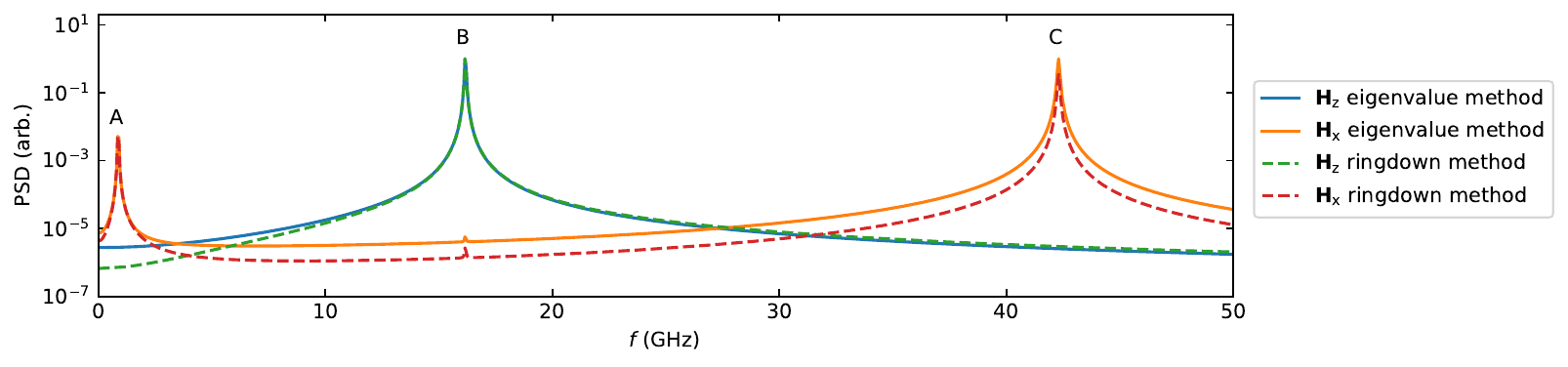}

  \textbf b

	\includegraphics[width=\columnwidth]{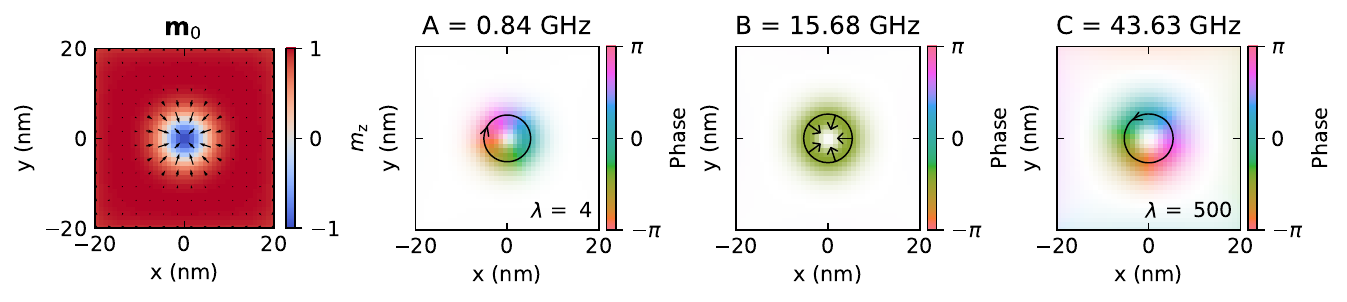}

	\textbf c
  \includegraphics[width=\columnwidth]{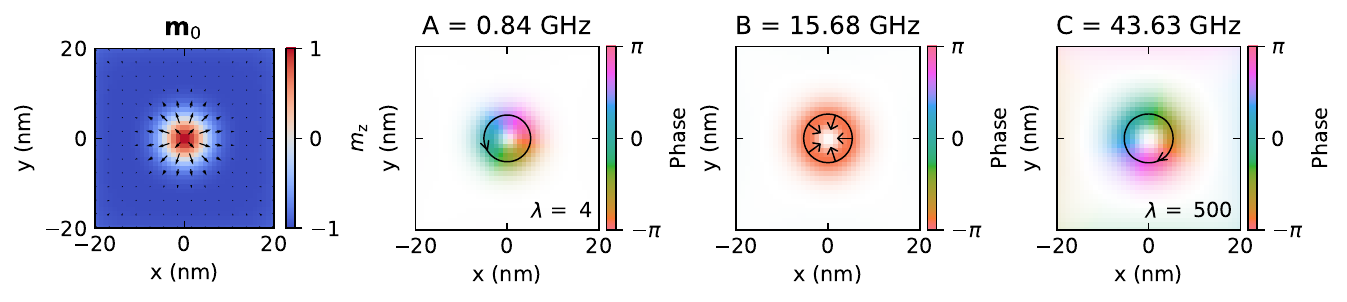}
	\caption{(a) Power spectral density (PSD) of Néel skyrmion in xy plane for two
    different excitations. An in-plane excitation field acting in the x
    direction excites modes A and C, an out-of-plane excitation field acting in
    the z direction excites mode~B. The solid lines show the spectra computed
    from the eigenvalue method, the dashed lines the spectra obtained from the
    ringdown method.
	  (b) Equilibrium configuration and first three eigenmodes~A,~B and~C of a Néel
    skyrmion with core pointing in the $-$z direction (Fig.~\ref{fig:geometries}a).
    (c) Equilibrium configuration and first three eigenmodes~A,~B, and~C of a Néel
    skyrmion with core pointing in the $+$z direction (Fig.~\ref{fig:geometries}b).
    \label{fig:single-skyrmion-fm}
  }
\end{figure}

Figure~\ref{fig:single-skyrmion-fm}a shows the power spectral density (PSD) for
a Néel skyrmion in a (chiral) ferromagnetic material (geometry
Fig.~\ref{fig:geometries}a, equilibrium configuration
shown in Fig.~\ref{fig:single-skyrmion-fm}b). The dashed lines show the PSD
obtained from the ringdown simulations. Solid lines show spectra computed from
the eigenvalues obtained with the eigenvalue method (see Methods for details how
the PSD is calculated).

The maxima of the PSD curves identify frequencies of normal modes, which we
label with letters. The computed PSD depends on the excitation signal. By
choosing a spatially uniform field (acting in the x or the z-direction), we have
chosen an excitation that can be realised experimentally. Modes~A (0.84$\,$GHz)
and C (43.63$\,$GHz) are excited by the in-plane excitation field, and mode~B
(15.68$\,$GHz) is excited by the out-of-plane excitation. The eigenvalue method
is numerically more efficient than the ringdown method, and also guaranteed to
find all eigenmodes (irrespective of the excitation), so we will use the
eigenvalue method for the rest of the paper.

Figure~\ref{fig:single-skyrmion-fm}b shows the spatial representation of the
three modes (see Supplementary Fig.~3 for more modes at higher
frequencies). Before we discuss the modes, we introduce the technique used to
visualise the normal modes: we employ a combined phase–amplitude mapping
approach and focus on the $m_{\mathrm{z}}$ component. At each spatial point $\mathbf{r}_i$ for a given frequency $f$,
amplitude and phase of the complex eigenvector are computed. The transparency of
each pixel is adjusted based on the amplitude. This method ensures that only
phase data from high-amplitude regions are prominently visible, reducing visual
clutter in low-signal areas. The phase is displayed using a cyclic colour map,
where hue represents the phase angle. Selecting a colormap with constant
lightness for all hues is crucial to not introduce artifacts in the
visualisation, more details on this can be found in Supplementary Fig.~1.

Mode~A is a clockwise (CW) gyrotropic mode. Mode~B is a breathing mode. Mode~C
is a counter clockwise (CCW) rotating mode where the skyrmion core position is
fixed and the surrounding skyrmion domain wall gyrates around the fixed core
position. Furthermore the background, in particular near the sample edges, is
also weakly excited.

To better visualise the type of motion we add overlays to the phase maps. For
the rotating modes we overlay the skyrmion trajectory (computed as first moment
of the topological charge density, details in Methods
Sec.~\ref{sec:data-analysis}). For better visibility, the trajectory is scaled
by an arbitrary factor~$\lambda$. An arrow marks the type of motion. The arrow
position indicates the skyrmion core position at a reference time, the arrow
direction provides a visual indication of the instantaneous direction of motion.
The different scaling values of $\lambda=4$ for modes~A and $\lambda=500$ for
mode~C show that the displacement of the skyrmion in mode~C is much smaller than
for mode~A. For the breathing mode, the overlayed circle contains the skyrmion
core and the arrows show expansion or contraction of the core at the reference
time. The reference time is arbitrary and only useful to see phase shifts in
different layers, but not suitable to directly compare different
eigenfrequencies or different systems.

In previous studies of eigenmodes of confined skyrmions both breathing modes
coupling to out-of-plane
excitation~\cite{kim2014,Mruczkiewicz2016,kim2017,beg2017a} and gyrotropic modes
coupling to in-plane excitation~\cite{Mruczkiewicz2016,kim2017,beg2017a} have
been reported, in agreement with the results here.

The low frequency gyrotropic mode~A can be described by the Thiele model, which
approximates the skyrmion magnetisation configuration as a rigid particle.
Within that model, which describes the skyrmion only through it's core position
and its gyrovector $\mathbf{G}$, one low frequency gyrotropic mode is expected
\cite{guslienko2017}. The sign of the gyrovector, which is proportional to
skyrmion topological charge~\cite{Guslienko2016}, determines the rotation
direction (CW or CCW) of that mode.

We repeat the study for the FM skyrmion but change the polarity of the skyrmion
so that the skyrmion core points in the opposite direction (now +z, geometry
Fig.~\ref{fig:geometries}b), which
reverses the sign of the topological charge. The results are shown in
Fig.~\ref{fig:single-skyrmion-fm}c. Indeed the rotation direction of the
gyrotropic mode~A has changed from CW to CCW and mode~C has changed from CCW to
CW.

\subsection{Skymion in SAF square geometry}\label{sec:one-skyrmion-saf}

\begin{figure}
  \sffamily
  \begin{minipage}{0.5\linewidth}
    \textbf a

    \includegraphics[width=\columnwidth]{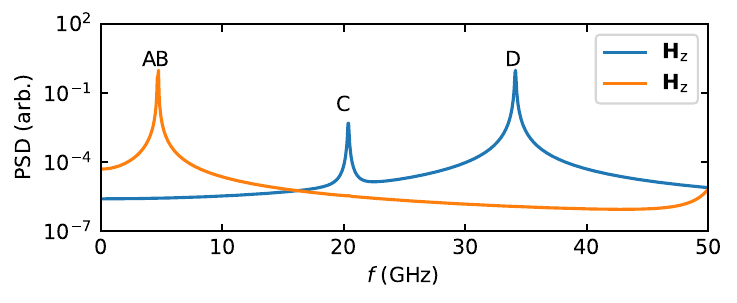}

  \end{minipage}%
  \begin{minipage}{.5\linewidth}
    \textbf b

    \includegraphics[width=\columnwidth]{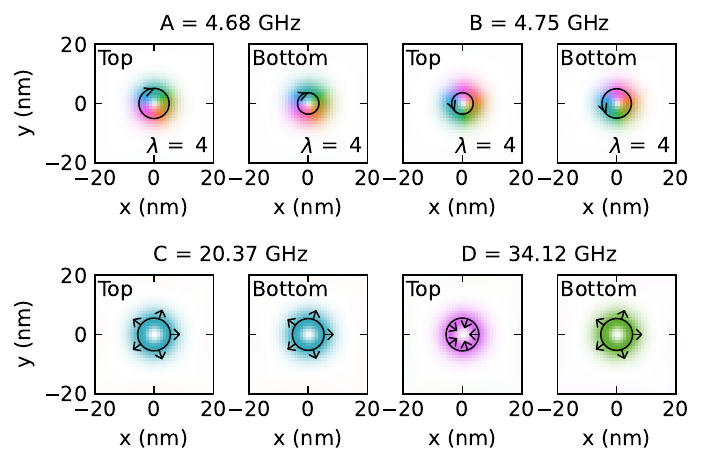}

  \end{minipage}

  \textbf c

  \includegraphics[width=.5\columnwidth]{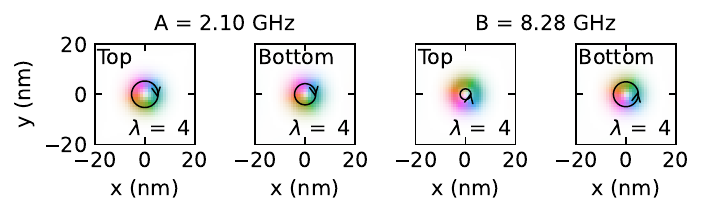}
	\caption{
    (a)~Power spectral density and (b)~normal modes for a SAF skyrmion (geometry
    as in Fig.~\ref{fig:geometries}c). The Néel skyrmion in the top layer has the
    vortex core pointing in $-$z direction, and the skyrmion in the bottom layer has
    the vortex core pointing in the +z direction. The layers are
    antiferromagnetically coupled. (a) The blue line shows the power spectral
    density for out-of-plane excitation acting in the z-direction. The orange line
    shows the power spectral density for an excitation acting in the in-plane
    x-direction. (b) The observed normal modes are gyrotropic (A~and~B) and
    breathing modes (C~and~D). (c)~Gyrotropic normal modes and frequencies for a SAF
    skyrmion in square geometry with top layer thickness of 2nm and bottom layer
    thickness of 1 nm ($L_\mathrm{x}=L_\mathrm{y}=40\,$nm) as shown in
    Fig.~\ref{fig:geometries}d.
    \label{fig:single-skyrmion-saf-40x40}
  }
\end{figure}

We now move to the skyrmions in a synthetic antiferromagnetic (SAF) bilayer. The
geometry is shown in Fig.~\ref{fig:geometries}c and described in the Methods
Sec.~\ref{sec:methods-geometry}, and hosts two antiferromagnetically coupled skyrmions:
one in the top layer with the core pointing in $-$z direction and another in the
bottom layer with the core pointing in +z direction.

Figure~\ref{fig:single-skyrmion-saf-40x40}a shows the power spectral densities
for the same in-plane and out-of-plane excitations as we use in
Sec.~\ref{sec:one-skyrmion-fm}, and the obtained normal modes up to 50\,GHz.
Modes~A and~B are gyrotropic modes, modes~C and~D are breathing modes. Higher
frequency modes start to hybridise with the sample, as we have already seen for
the ferromagnetic skyrmion (mode~C in Fig.~\ref{fig:single-skyrmion-fm}b and~c),
and are shown in Supplementary Fig.~6.

\subsubsection{Gyrotropic modes}\label{section:skyrmion_40x40_saf_gyrotropic}

The first two modes~A and~B in Fig.~\ref{fig:single-skyrmion-saf-40x40} are
gyrotropic modes with very similar frequencies $f_\mathrm{A} = 4.68\,$GHz and
$f_\mathrm{B} = 4.75\,$GHz and are nearly indistinguishable in the spectrum.

As for the corresponding ferromagnetic single layer study, the gyrotropic modes
are excited by the in-plane field. The two skyrmions, one in the top and one in
the bottom layer of the SAF, gyrate in the same orientation: clockwise for
mode~A and counter clockwise for mode~B.

From the study of the ferromagnetic system (Sec.~\ref{sec:one-skyrmion-fm}) we
know that given a skyrmion core polarisation, one rotation sense is preferred.
In the SAF setup, the skyrmion in the top layer has core polarity $-1$ and, due
to the AFM coupling, the skyrmion in the bottom layer has core polarity $+1$.
Due to the interlayer coupling forces, top and bottom layer rotate (mostly)
coupled. The system of the two skyrmions in the SAF sample is frustrated: the
preferred rotation direction for the skyrmion in the top layer is the opposite
of the preferred direction for the skyrmion in the bottom layer.

The frustration also manifests in the skyrmion trajectories in
Fig.~\ref{fig:single-skyrmion-saf-40x40}. The trajectory of the top layer
skyrmion for mode~A shows a larger diametre than the trajectory for the bottom
layer skyrmion. For mode~B the trajectory of the bottom layer skyrmion has the
larger diametre. This is in line with the core down polarisation of the top
layer skyrmion preferring the CW rotation, and the core up polarisation in the
bottom layer skyrmion preferring the CCW rotation
(Sec.~\ref{sec:one-skyrmion-fm}). Due to the interlayer coupling, both layers
show the same rotation but the non-preferred rotation direction has reduced
amplitude (i.e. reduced trajectory diametre).

Figure~\ref{fig:single-skyrmion-saf-40x40}c shows the gyrotropic normal modes
for a modified SAF geometry with non-identical layer thicknesses: the top layer
has an increased thickness of 2\,nm, while the bottom layer is 1\,nm thick
(geometry Fig.~\ref{fig:geometries}d). In
the symmetric case (both layers 1\,nm), the eigenmodes at 4.68\,GHz and 4.75\,GHz are
nearly degenerate. In the asymmetric geometry, this near-degeneracy is lifted,
and the modes split into mode~A at 2.1\,GHz and mode~B at 8.28\,GHz. Mode~A
corresponds to a CW gyrotropic motion, which is the preferred rotation direction
in the top layer. Since the top layer accounts for approximately two-thirds of
the total magnetic volume, this lowers the overall mode frequency. Conversely,
the CCW mode is unfavourable in the thicker top layer, leading to an increase in
its energy and thus a higher frequency. This behaviour is consistent with our
interpretation of the coupled modes (decreased amplitude for the non-preferred
rotation sense). Finally, we note that combining layers of different thicknesses
can be advantageous in applications where a finite stray field is required.

A similar splitting of frequencies for clockwise and anticlockwise gyrotropic
modes, where the skyrmions in the top and bottom layers rotate in the same
direction, has been reported for a skyrmion in a SAF
confined to a disk geometry, Ref.~\cite{xing2018}, Fig.~4, 61\,GHz and 66\,GHz.
However, the behaviour reported for the other modes is not consistent with our
findings, and the overall interpretation of the remaining modes differs from
what we observe. To the best of our knowledge, no other studies have reported a
comparable splitting of these modes in SAF skyrmions.

\subsubsection{Breathing modes}\label{section:skyrmion_40x40_saf_breathing}

Modes~C and~D in Fig.~\ref{fig:single-skyrmion-saf-40x40}b are breathing modes,
in which the skyrmion radius grows and shrinks
periodically~\cite{kim2017,lonsky2020a}. The visualisations show that in mode~C
at $f_\mathrm{C}=20.37\,$GHz the top and bottom layer breath in phase, whereas
in mode D at $f_\mathrm{D}=34.12\,$GHz the top and bottom layer breath out of
phase. The energetic impact of out-of-phase behaviour between top and bottom
layer is high due to the large interface area over which the RKKY coupling is
active. These in-phase and out-of-phase breathing modes are consistent with the
results~\cite{lonsky2020a} reported for a skyrmion in a SAF that is confined by
the disk-like geometry.

\subsection{Skyrmion in SAF rectangle geometry: translational modes}

\begin{figure}
  \begin{minipage}{0.5\linewidth}
    \sffamily \textbf a
    \includegraphics[width=\linewidth]{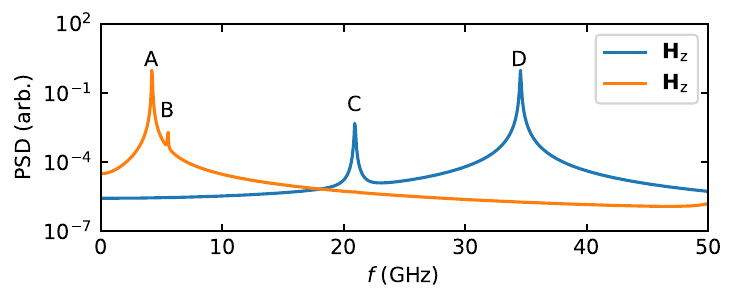}
  \end{minipage}%
  \begin{minipage}{0.5\linewidth}
    \sffamily \textbf b
    \includegraphics[width=\linewidth]{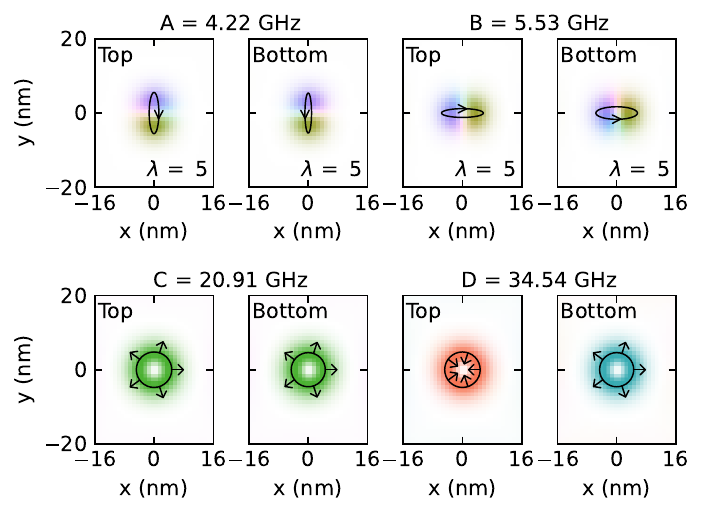}
  \end{minipage}

  \caption{(a) PSD and (b) lowest frequency normal modes of a skyrmion in
    a SAF with sample dimensions $L_\mathrm{x}=32\,$nm and
    $L_\mathrm{y}=40$\,nm. We classify modes A and B as translational
    modes, and modes C and D as breathing modes. For more details refer to the
    main text.
    \label{fig:single-skyrmion-saf-40x32}
  }
\end{figure}

Figure~\ref{fig:single-skyrmion-saf-40x32} shows the PSD and the normal modes of
the SAF skyrmion in a sample with $L_\mathrm{x}=32\,$nm and
$L_\mathrm{y}=40$\,nm (geometry in Fig.~\ref{fig:geometries}e). Apart from the
dimension in the x-direction, this is the same system as shown in
Fig.~\ref{fig:single-skyrmion-saf-40x40} where $L_\mathrm{x}=L_\mathrm{y}=40$.
The breathing modes C and D are very similar in
Fig.~\ref{fig:single-skyrmion-saf-40x32}b and
Fig.~\ref{fig:single-skyrmion-saf-40x40}b, and the small change in frequency is
probably related to the changed confinement potential by going from
$L_\mathrm{x}=40$\,nm to $L_\mathrm{x}=32$\,nm.

However, modes A and B differ significantly. For the square geometry
(Fig.~\ref{fig:single-skyrmion-saf-40x40}b), we find two gyrotropic modes A and
B, where for mode~A top and bottom layer both show CW gyration. For the
rectangular geometry (Fig.~\ref{fig:single-skyrmion-saf-40x32}b) we find that
mode~A exhibits a CW rotation in the top layer and a CCW rotation in the bottom
layer, i.e. the skyrmion in each layer gyrates in its preferred direction. The
skyrmion core positions describe elliptic trajectories in each layers (although
with opposite gyration sense). The SAF skyrmion core position (the average of
the top and bottom layer skyrmion position) describes a (nearly) linear motion
along the y-direction as the x-components of the elliptic motion in the two
layers cancel each other due to the opposite rotation sense. For the lower
frequency mode~A, this translational motion occurs in the y-direction, along the
long axis of the sample, and for the higher frequency mode~B a corresponding
motion occurs in the x-direction, along the short axis of the sample.

\begin{figure}
  \sffamily
  \begin{minipage}{0.5\linewidth}
    \textbf a
    \includegraphics[width=\linewidth]{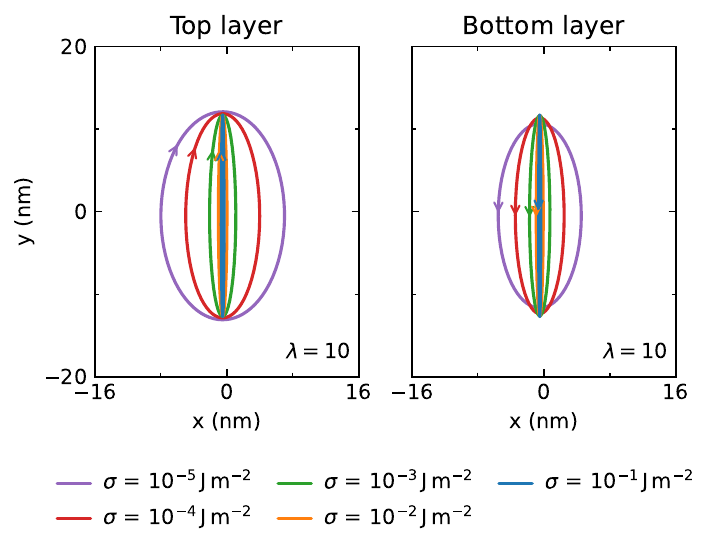}
  \end{minipage}%
  \begin{minipage}{0.5\linewidth}
    \textbf b
    \includegraphics[width=\linewidth]{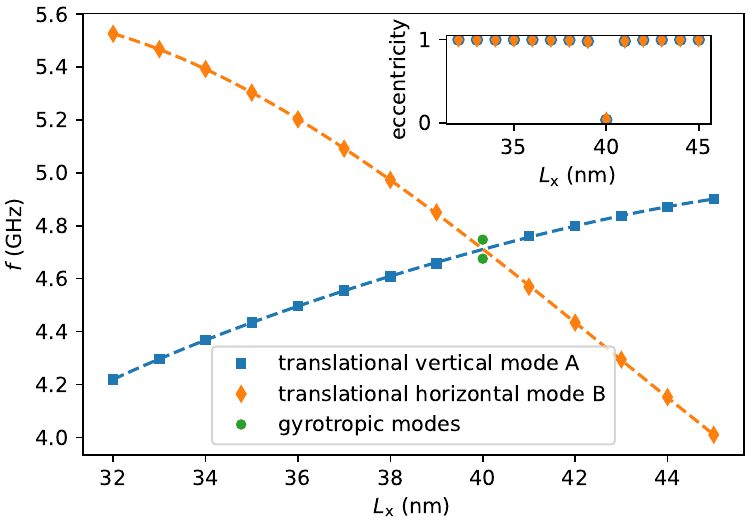}
  \end{minipage}

  \caption{ (a) Skyrmion core trajectories for mode~A of SAF skyrmion in 32\,nm
    $\times$ 40\,nm geometry for different interlayer coupling
    strengths~$\sigma$. (b) Frequency of the translational modes~A and~B (see
    Fig.~\ref{fig:single-skyrmion-saf-40x32}) as a function of geometry
    $L_\mathrm{x}$. All studied systems have $L_\mathrm{y} = 40\,$nm. For the
    quadratic sample with $L_\mathrm{x} =40$\,nm and $L_\mathrm{y} = 40$\,nm, we
    find two gyrotropic modes, where skyrmion cores moves on a circular
    trajectory (see Fig.~\ref{fig:single-skyrmion-saf-40x40}b). The inset shows
    the eccentricity of the ellipse of the SAF skyrmion trajectories (summed
    trajectories of top and bottom layer skyrmion). We see a sharp transition
    from just below 1 (ellipse) for $L_\mathrm{x}\neq L_\mathrm{y}$ to 0
    (circle) for $L_\mathrm{x}=L_\mathrm{y}$.
      \label{fig:single-skyrmion-saf-scaling-behaviour}
    }
\end{figure}

Figure~\ref{fig:single-skyrmion-saf-scaling-behaviour}a shows the trajectory of
the skyrmion in the top and bottom layer for the translational mode~A for a
number of different interlayer coupling strengths $\sigma$. The weaker coupling
values show that the motion is CW in the top layer and CCW in the bottom layer.
As the coupling strength $\sigma$ increases, the elliptical trajectory becomes
increasingly more elongated and narrower. For the strongest values studied
$\sigma = 0.1\,$Jm$^{-2}$ the trajectory (in both layers) appears as a line
(within the resolution of our figure). We interpret this as the interlayer
coupling strength enforcing increasingly more identical spin rotations in the
top and bottom layer as $\sigma$ increases. The x-components of the top layer
skyrmion core position grows positively when the bottom layer skyrmion core
position grows negatively within one rotation cycle, whereas the y-positions of
both layers change in-phase. As the coupling increases towards infinity, the
out-of-phase motion is suppressed and only translational motion in the
y-direction remains.

We have studied translational modes~A and~B systematically with $L_\mathrm{x}$
changing from from 32\,nm to 45\,nm in steps of 1\,nm (the mesh discretisation
is 1\,nm) and keeping $L_\mathrm{y}=40\,$nm fixed.
Figure~\ref{fig:single-skyrmion-saf-scaling-behaviour}b shows the frequency of
the vertical translational mode~A (blue squares) and the horizontal
translational mode~B (orange diamonds) as a function of $L_\mathrm{x}$. The
dashed line is a 4-th order polynomial fit through the data points. The mode
aligned with the longer sample axis has the lower frequency, i.e. mode~A for
$L_\mathrm{x} < L_\mathrm{y}$ and mode~B for $L_\mathrm{x} > L_\mathrm{y}$.
Increasing the sample aspect ratio increases the mode splitting. The two green
dots at $L_\mathrm{x}=40\,$nm are the two gyrotropic modes (mode~A and~B in
Fig.~\ref{fig:single-skyrmion-saf-40x40}b) observed in the square geometry.

From the eccentricity of the SAF skyrmion core trajectory (inset in
Fig.~\ref{fig:single-skyrmion-saf-scaling-behaviour}b), we find a circular
trajectory only for $L_\mathrm{x}=L_\mathrm{y}=40\,$nm. For this square
geometry, we call modes A and B gyrotropic modes. For aspect ratios
$L_\mathrm{x}/L_\mathrm{y} \ne 1$, we call modes A and B translational modes.
The names \emph{gyrotropic} and \emph{translational} reference the trajectory of
the SAF skyrmion core.

The behaviour of the two modes near $L_\mathrm{x} = L_\mathrm{y} = 40\,$ nm (green dots in Fig.~\ref{fig:single-skyrmion-saf-scaling-behaviour}b) is
characteristic of an avoided crossing. As $L_\mathrm{x}$ is varied through the square
geometry, the vertical translational mode~A and the horizontal translational
mode~B do not intersect but instead repel each other, with their frequencies
splitting as $\omega = \omega_0 \pm g$ around a common frequency $\omega_0$. This hybridization arises
because the two modes are not independent: the interlayer coupling mixes the
elliptical layer trajectories, so that near the symmetric point the eigenmodes
are no longer purely ``vertical'' and ``horizontal'' translations but symmetric and
antisymmetric superpositions of the two. At exactly $L_\mathrm{x} = L_\mathrm{y}$ the geometric
anisotropy that distinguishes the two translation directions vanishes, the two
hybridized modes become degenerate in trajectory shape (circular, eccentricity →
0, see inset of Fig.~\ref{fig:single-skyrmion-saf-scaling-behaviour}b), and they recover the gyrotropic character described in
Sec.~\ref{sec:one-skyrmion-saf}. Away from the square point, increasing $|L_\mathrm{x} - L_\mathrm{y}|$ restores the
distinct vertical and horizontal translational identities of the two branches.

In summary, we have investigated a single skyrmion in a confined geometry in a
ferromagnetic and synthetic antiferromagnetic material. For the ferromagnetic
material, a square geometry, and a skyrmion with core polarity $-1$ (core
pointing into $-$z direction), we find three eigenmodes: gyrotropic (CW),
breathing and rotational (CCW) (Fig.~\ref{fig:single-skyrmion-fm}). One of the
gyration directions is preferred, and the preferred direction is linked to the
skyrmion core polarity. In the synthetic antiferromagnetic in a square geometry
the two skyrmions are coupled through the interlayer RKKY. We find two nearly
degenerated gyrotropic modes where the gyration direction in top and bottom
layers is identical within each mode (Fig.~\ref{fig:single-skyrmion-saf-40x40}).
At higher frequencies, breathing modes are found, which are in-phase and
out-of-phase (across the layer boundary), respectively. Once the symmetry of
$L_\mathrm{x}=L_\mathrm{y}$ in the square geometry is broken so that
$L_\mathrm{x} \ne L_\mathrm{y}$, the two gyrotropic modes disappear and are
replaced by translational modes (Fig.~\ref{fig:single-skyrmion-saf-40x32}),
where the gyration direction in top and bottom layer is opposite, resulting in
effectively linear motion of the skyrmion cores.

\subsection{Two skyrmions}\label{sec:results-two-skyrmions}

\begin{figure}
  \sffamily
  \begin{minipage}{0.5\linewidth}
    \textbf a
    \includegraphics[width=\linewidth]{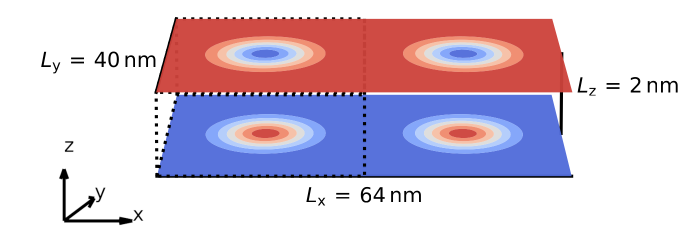}
    \textbf b
    \includegraphics[width=\linewidth]{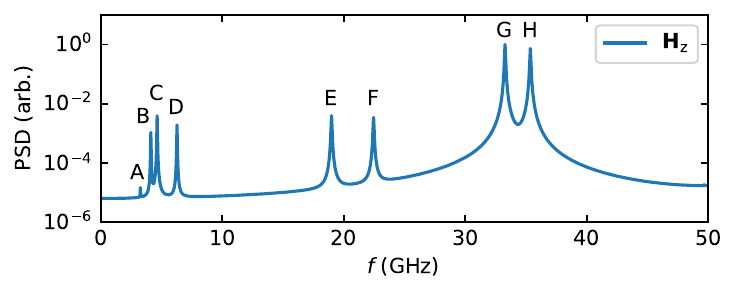}
  \end{minipage}%
  \begin{minipage}{.5\linewidth}
    \textbf e
    \includegraphics[width=\linewidth]{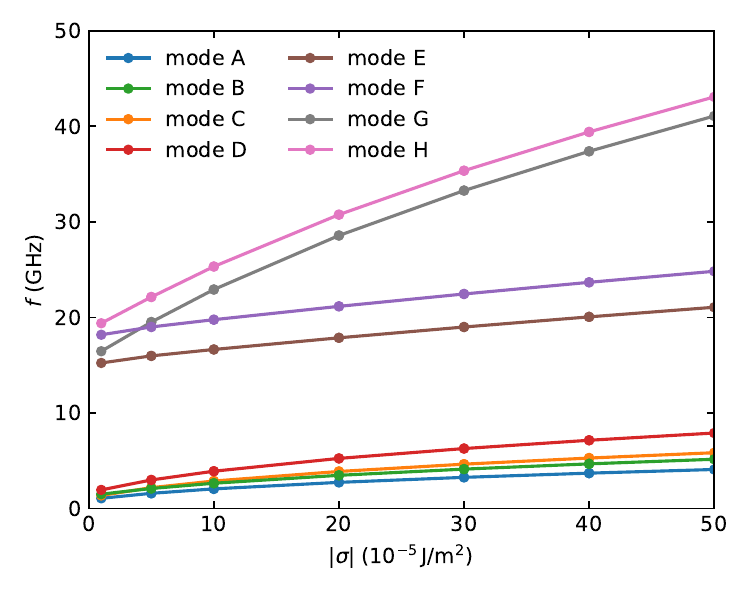}
  \end{minipage}

  \begin{minipage}{0.5\linewidth}
    \textbf c
    \includegraphics[width=\columnwidth]{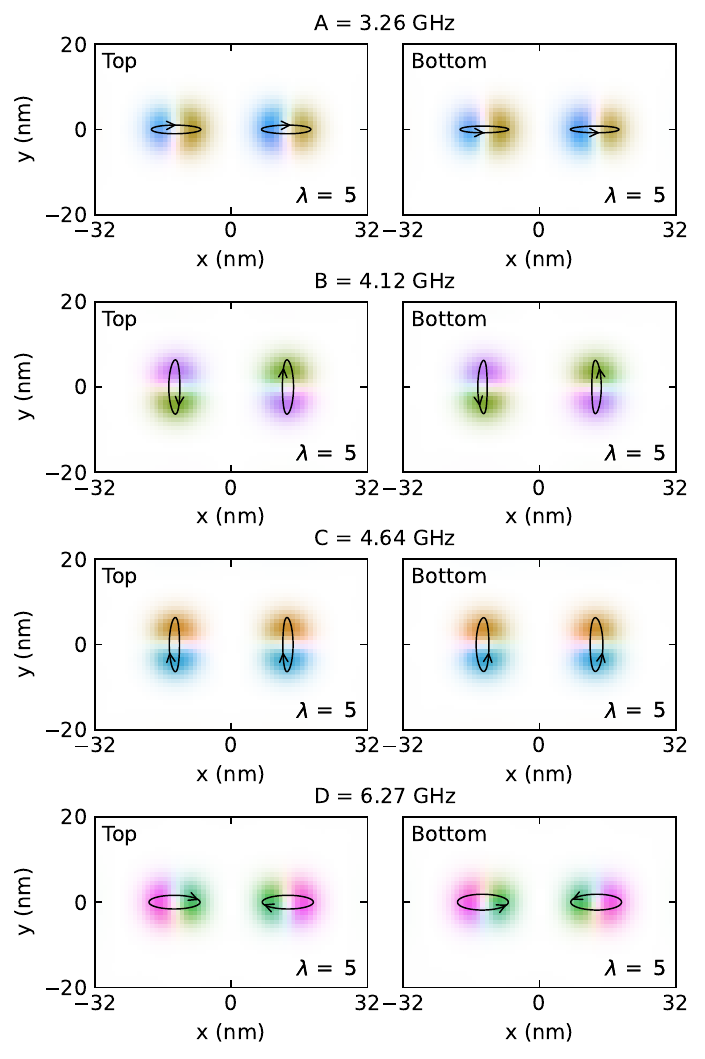}
  \end{minipage}%
  \begin{minipage}{0.5\linewidth}
    \textbf d
    \includegraphics[width=\columnwidth]{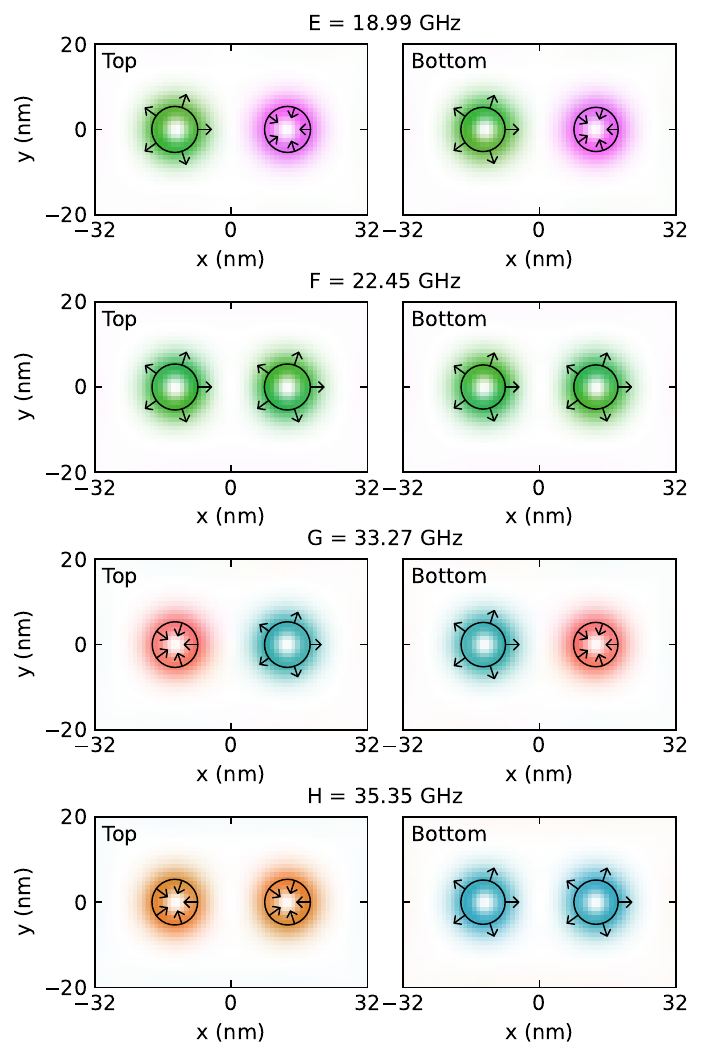}
  \end{minipage}

	\caption{ (a) Schematic of the two skyrmion system confined in a $(64 \times
    40 \times 2)\,\mathrm{nm}^3$ geometry. (b) PSD for the two skyrmion system.
    The excitation field is acting in the out-of-plane direction and only
    excites the left SAF skyrmion, indicated by the dotted lines in subfigure~a.
    (c) Normal modes~A to~D.
    All four modes have translational character. (d) Normal modes~E to~H. All
    four modes are breathing modes.
    (e) Frequencies of the first eight eigenmodes for
    varying interlayer RKKY coupling strength $|\sigma|$. For weak coupling we
    observe a different order of the modes.
    \label{fig:two-skyrmions-saf}
  }
\end{figure}

Next, we study a strip of length $L_\mathrm{x}=64\,$nm, which can hold two
skyrmions in each layer. Figure~\ref{fig:two-skyrmions-saf}a illustrates the
geometry (also shown in Fig.~\ref{fig:geometries}f). We relax the system into two pairs of N\'eel type skyrmions, which are
antiferromagnetically coupled between top and bottom layer with a core-to-core
spacing of 32$\,$nm within a given layer.

For the spectrum calculation, the excitation field is applied to the left pair
of top-bottom skyrmions only (i.e. for $x < 32\,$nm, in
Fig.~\ref{fig:two-skyrmions-saf}a indicated by the dotted lines) and applied in
the out-of-plane direction. For subsequent studies with more than two skyrmions,
we will also excite only the left-most SAF skyrmion.

The PSD shown in Fig.~\ref{fig:two-skyrmions-saf}b, shows eight normal modes. We
will identify below that the first four modes~A to~D are translational modes,
and modes~E to~H are breathing modes.

Figure~\ref{fig:two-skyrmions-saf}c shows phase-amplitude plots of the normal
modes~A to~D. We see elliptical trajectories for all 4 modes, with opposite
gyration sense in top and bottom layer for each mode. This corresponds to the
translation modes we identified for the single SAF skyrmion
(Fig.~\ref{fig:single-skyrmion-saf-40x32}) but here extended to two SAF
skyrmions. The lowest frequency mode~A is now showing horizontal translation
presumably because the shape anisotropy of the sample make the x-direction an
easy axis for skyrmion displacement.

The same geometry ($L_\mathrm{x}=64\,$nm and $L_\mathrm{y}=40\,$nm) with two
skyrmions has been studied as a single-layer ferromagnetic system \cite{kim2017}
and the lowest frequency mode found shows elliptical trajectories with the long
axis along the x-direction (resembling qualitatively the orientation of the
elliptical trajectories of the skyrmion cores we find in the top layer in
mode~A). This supports the interpretation that an elliptical shape of the
trajectory is independent from the interlayer coupling and driven by the shape
anisotropy of the sample (see also Supplementary Fig.~5). However, the interlayer coupling makes the elliptical
trajectories more anisotropic as the two layers favour opposite gyration
directions. The skyrmion pairs forming a SAF skyrmion are always in-phase along
the long axis of the ellipse between the RKKY coupled layers.

As we have two SAF skyrmions next to each other, they can show in-phase or
out-of-phase motion relativ to the other SAF skyrmion: modes~A and~C are
in-phase between the neighbouring skyrmions and~B and~D are out-of-phase.

The amplitude phase maps for modes~E, F, G and~H are shown in
Fig.~\ref{fig:two-skyrmions-saf}d. These are all breathing modes. The lowest
frequency breathing mode~E shows in-phase dynamics for the top and bottom
skyrmions, but out-of-phase coupling between the left and the right skyrmion.
While the left SAF skyrmion contracts, the right SAF skyrmion undergoes radial
expansion, corresponding to a phase shift of $\pi$ between the left and right
SAF skyrmion. The out-of-phase dynamics of skyrmions in a given layer provides
an opportunity for one skyrmion to grow and fill the space when the other
shrinks, thus reducing the energic cost of this mode. At the same time, the
interlayer coupling is strong enough to reward in-phase dynamics between the top
and bottom layer.

Mode~F shows in-phase behaviour across the layers, and between left and right
skyrmion. The radial expansion and contraction occur simultaneously for the left
and right skyrmions, and in both layers. We hypothesize that mode~F has a higher
energy than mode~E because the left and right skyrmion compete for space when
expanding simultaneously.

Mode~G decouples motion of top and bottom layer: when the skyrmion in the top
layer grows, the corresponding one in the bottom layer shrinks. This is combined
with an out-of-phase behaviour between left and right skyrmion.

The most energetically expensive combination is seen for mode~H: out-of-phase
across the layer interface, and in-phase between left and right skyrmion.

The dependence of the resonance frequencies of the first eight modes on the
interlayer coupling strength $\lvert\sigma\rvert$ is shown in
Fig.~\ref{fig:two-skyrmions-saf}e. Overall, we can see a decrease in frequency
with decreasing interlayer coupling. This is expected as the total energy of the
system decreases with decreasing~$|\sigma|$. For the two modes~G and~H we can
observe a stronger frequency decrease with decreasing~$|\sigma|$ than for the
other modes, and for $|\sigma| \rightarrow 0$ they approach modes~E and ~F,
respectively. In the limit $\sigma = 0$ the two layers are uncoupled (ignoring
demagnetisation effects for a moment), and the four eigenfrequencies collapse to
just two, with the modes~E and~G (left right out of phase) at the lower
frequency, and modes~F and~H (left right in phase) at the higher frequency. This
mode reordering (swapping of modes~F and~G) can be seen just below $|\sigma| = 5
\cdot 10^{-5}\,\mathrm{J m}^{-2}$, where modes~F and~G cross. In reality the two
layers are for $|\sigma| = 0$ still weakly coupled via the demagnetisation field
and the frequency degeneracy at $\sigma=0$ is lifted (instead further mode
re-ordering would occur for $|\sigma| > 0$). For better clarity we exclude
results for $|\sigma| < 1 \cdot 10^{-5}\,\mathrm{J m}^{-2}$ in
Fig.~\ref{fig:two-skyrmions-saf}e, so that RKKY is dominating the interlayer
coupling for everything shown in the figure. In most parts of this study we use
the value $|\sigma| = 30 \cdot 10^{-5}\,\mathrm{J\,m^{-2}}$.

In summary, for the two SAF skyrmion system, we have established translational
modes in the frequency range 3 to 6$\,$GHz (Fig.~\ref{fig:two-skyrmions-saf}c)
and breathing modes in the frequency range 18 to 35$\,$GHz
(Fig.~\ref{fig:two-skyrmions-saf}d). For different values of $\sigma$, the
frequencies adjust: stronger coupling increases the frequency of the modes.

\subsection{Five skyrmions}\label{sec:five-skyrmions}

\begin{figure}
  \sffamily
  \begin{minipage}{0.5\linewidth}
    \textbf a
    \includegraphics[width=\linewidth]{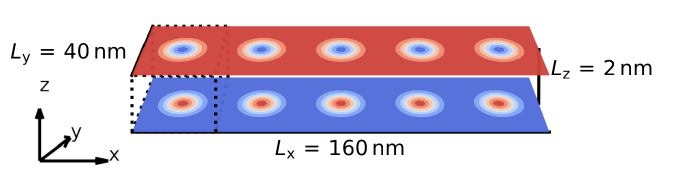}
  \end{minipage}%
  \begin{minipage}{0.5\linewidth}
    \textbf b
    \includegraphics[width=\linewidth]{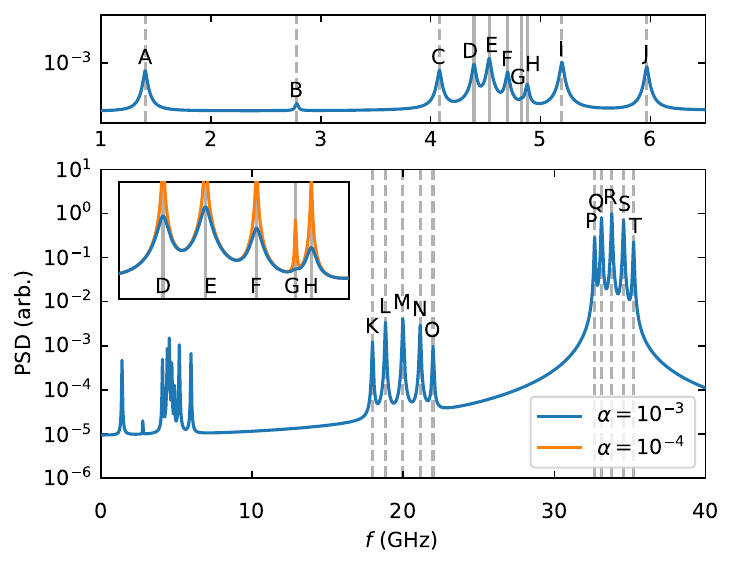}
  \end{minipage}

  \begin{minipage}{0.5\linewidth}
    \textbf c
    \includegraphics[width=\linewidth]{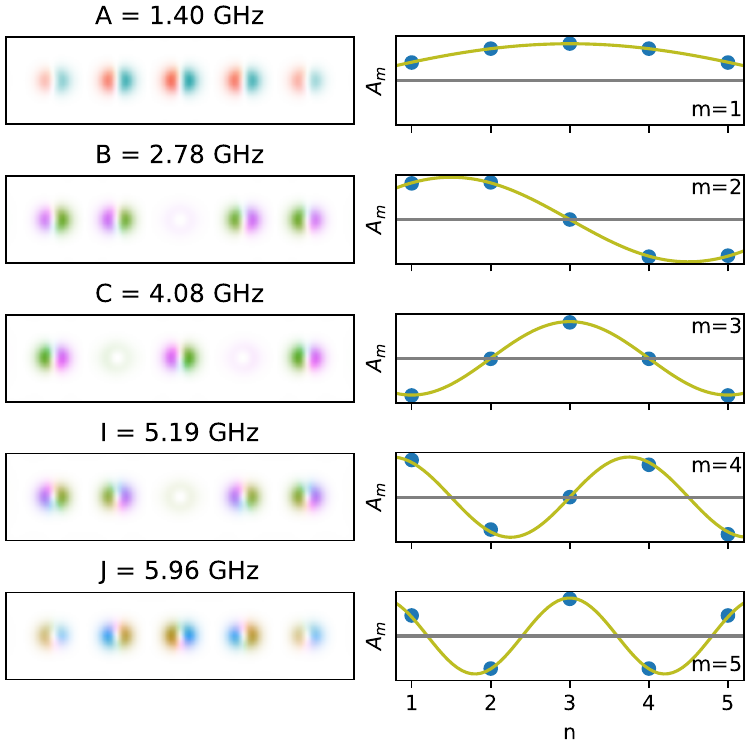}
  \end{minipage}%
  \begin{minipage}{0.5\linewidth}
    \textbf d
    \includegraphics[width=\linewidth]{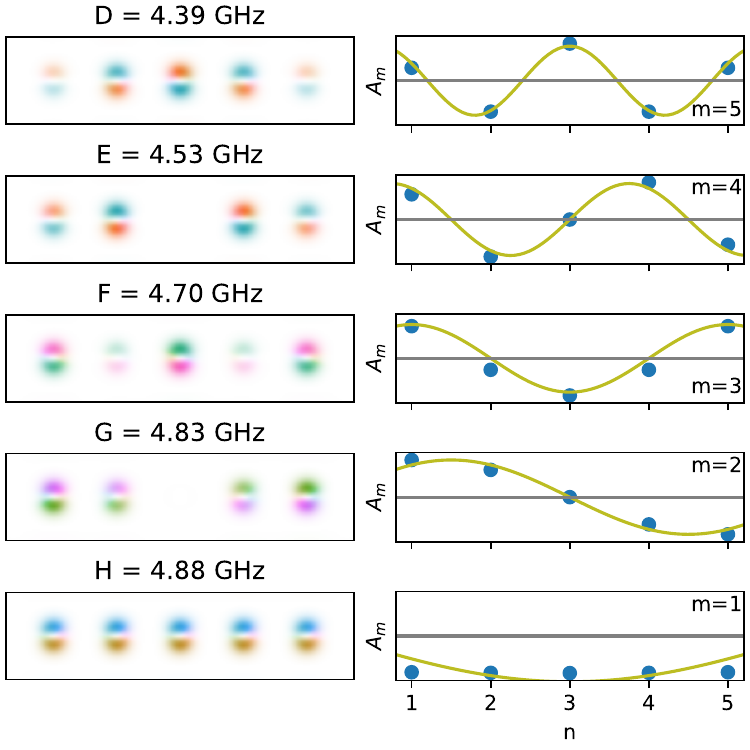}
  \end{minipage}

  \caption[PSD]{(a) System geometry for the five skyrmion study. The dotted lines
    indicate the region where the excitation pulse is applied. (b) PSD in response to an out-of-plane excitation.
    The low-frequency range up to
    6\,GHz is shown in more detail in the top panel. The vertical lines indicate the eigen
    frequencies. As a guide to the eye, dashed and solid lines in the top panel indicate translational
    horizontal and vertical modes, respectively. The inset shows the
    translational vertical modes for two different
    damping values.
    (c) Left: Phase amplitude maps for the translation horizontal modes.
    Right:
    Amplitudes $A_m(n)$ of collective skyrmion coordinates from
    simulation (blue points) and amplitudes $\tilde{A}_m(n)$
    (\ref{eq:coupled-oscillator-amplitude}) of the associated standing
    mode $m$ (green line).
    (d) Phase amplitude maps and amplitudes of the collective skyrmion coordinates for the translational
    vertical modes.
    \label{fig:five-skyrmions-saf}
  }
\end{figure}

We extend the analysis of the magnetisation modes to a five-skyrmion strip
geometry of length $L_\mathrm{x} = 160$\,nm as shown in
Fig.~\ref{fig:five-skyrmions-saf}a and Fig.~\ref{fig:geometries}g.

Figure~\ref{fig:five-skyrmions-saf}b shows the PSD obtained for the given
excitation pulse (out-of-plane on the area occupied by the left-most SAF
skyrmion). For extra clarity the eigenfrequencies are also marked with lines. We
note that mode~B couples only very weakly to the chosen excitation and modes~G
and~H have very nearby frequencies leading to overlapping peaks for the standard
damping $\alpha=10^{-3}$ used throughout the work.

The inset in Fig.~\ref{fig:five-skyrmions-saf}b shows a smaller frequency range
containing modes~D to~H for two different damping values. Here we can more
clearly see that modes~G and~H cannot fully be resolved with $\alpha=10^{-3}$,
but split into two distinct peaks for $\alpha=10^{-4}$.

Modes~A to~J are translation modes (in Fig.~\ref{fig:five-skyrmions-saf}b
horizontal and vertical translation is indicated by dashed and solid lines,
respectively), modes~K to~O are breathing modes where top and bottom layer are
in phase and modes~P to~T are breathing modes where top and bottom layer are out
of phase.

For five skyrmions we see that the different types of modes start to form bands.
For the breathing modes all eigenmodes in a band have relatively similar
intensity. These modes all couple well to the out-of-plane excitation. Coupling
efficiency to the translational modes depends stronger on the specifics of the
mode. Furthermore, horizontal and vertical translational modes have overlapping
frequencies, making the spectrum appear less clean.

The phase amplitude maps for the translational horizontal modes are shown on the
left in Fig.~\ref{fig:five-skyrmions-saf}c, those for the translational vertical
modes on the left in Fig.~\ref{fig:five-skyrmions-saf}d. For the breathing modes
refer to Supplementary Fig.~9.

The lowest frequency mode~A for the 5 skyrmion system
(Fig.~\ref{fig:five-skyrmions-saf}c top) exhibits horizontal translation of the
skyrmion cores. All 5 skyrmions move in-phase. The amplitude of the excitations
varies from skyrmion to skyrmion. It is smallest for skyrmions~1 (on the left)
and~5 (on the right) and largest for skyrmion~3 (in the centre). The lower
amplitude can be seen through the lower opacity of the colours in the amplitude
phase map. Additionally, the plot to the right in
Fig.~\ref{fig:five-skyrmions-saf}c top, which shows the amplitude of skyrmion
centre displacement for individual skyrmions, emphasises this difference. The
second lowest frequency mode~B also shows variation of excitation amplitude: the
centre skyrmion 3 is not significantly excited and thus appears (virtually)
transparent in the phase-amplitude plot. The two skyrmions to the left (1 and 2)
move together (i.e. in-phase) but are out-of-phase relative to the two skyrmions
on the right (4 and 5). Similar inhomogeneities in the intensity distribution
can also be observed for most of the other translational modes shown in
Figs.~\ref{fig:five-skyrmions-saf}c and~\ref{fig:five-skyrmions-saf}d.

A model has been proposed for gyrotropic modes of skyrmions in a
ferromagnet~\cite{kim2017} and for breathing modes of skyrmions in a
ferromagnet~\cite{kim2018a} where the amplitude $A$ of the skyrmion excitation
follows a standing wave
\begin{equation}
  \tilde{A}_{m}(n) = a \sin(k_m n),
  \label{eq:coupled-oscillator-amplitude}
\end{equation}
with $k_m = m\pi / (N+1)$, where $n=1, 2, \ldots, N$ is the index of the
skyrmion ranging from 1 (left-most skyrmion) to $N=5$ (right-most skyrmion),
$m=1, 2, \ldots, N$ denotes the standing wave mode index, and $a$ is a fitting
constant. Through (\ref{eq:coupled-oscillator-amplitude}) the boundary
conditions are that (virtual) skyrmions at positions $n=0$ and $n=N+1$ have an
amplitude of~0.

Can the same model be applied for the more complex two-layer SAF system under
investigation here? Figure~\ref{fig:five-skyrmions-saf}c shows good fits for the
translational horizontal modes. We find that the model also describes well the
breathing modes~K to~T (see Supplementary Fig.~2). However, the model seems
not appropriate for the translational vertical modes as shown in
Fig.~\ref{fig:five-skyrmions-saf}d. In particular mode~H at 4.88\,GHz shows
uniform translational motion of all 5 skyrmion cores: it is impossible to fit a
sin function as in (\ref{eq:coupled-oscillator-amplitude}) with amplitude $
\tilde{A}_m(0) = 0$ and $\tilde{A}_m(N+1) = 0$.

In summary, for five skyrmions, we find the same translational and breathing
modes as for one and two skyrmions. The different types of modes start to form
bands, where for a given mode type (e.g. breathing top bottom in-phase) the
extrema are out-of-phase and in-phase excitation of neighbouring SAF skyrmions,
respectively.

\subsection{Signal propagation}\label{sec:signal-propagation}

\begin{figure}
  \includegraphics[width=\linewidth]{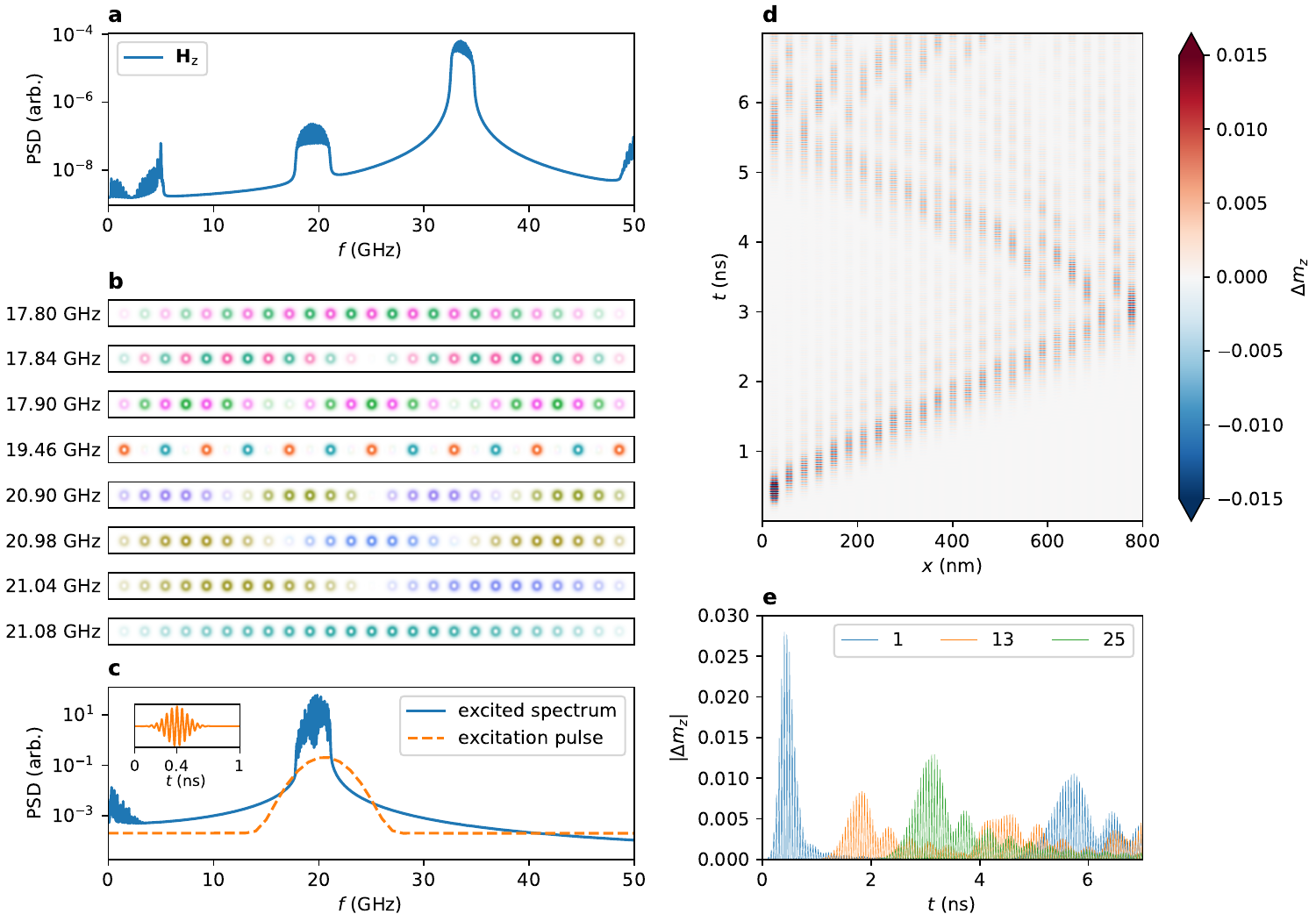}
  \caption{(a) PSD for 25 skyrmions. (b) Subset of the breathing modes in the 20\,GHz
    band. (c) Spectrum after exciting the 20\,GHz band with a gaussian pulse. Only the left-most
    skyrmion is excited by the pulse. The pulse power distribution in frequency
    space is shown with the dashed line. Inset: shape of the excitation pulse in
    time domain.
    (d) Propagation of excitation through the strip. (e) Excitation of selected skyrmions in
    the strip (1: left-most skyrmion, 13: centre skyrmion, 25: right-most
    skyrmion). The applied excitation has its maximum at $t=0.4$\,ns.}
  \label{fig:25-skyrmions}
\end{figure}

Finally, we show results for a longer chain of 25 SAF skyrmions (geometry Fig.~\ref{fig:geometries}h), in the spirit
of similar studies done for a ferromagnetic chain of skyrmions
\cite{kim2017,kim2018a}. In this system we can study signal propagation along
the skyrmion chain.

Figure~\ref{fig:25-skyrmions}a shows the power spectrum for out-of-plane
excitation of the left-most SAF skyrmion. The different types of eigenmodes
(translation, breathing) now have many possible realisations, resulting in the
formation of distinct bands in the PSD. At low frequencies, we can see the
translational modes, horizontal and vertical translation are intermixed as we
have already seen for five skyrmions. At around 20\,GHz we see a band of top
bottom in-phase breathing modes, at 35\,GHz top bottom out-of-phase breathing
modes. Near 50\,GHz we see the onset of a higher rotational mode, similar to
mode~C in the ferromagnetic study (Fig.~\ref{fig:single-skyrmion-fm}).

Figure~\ref{fig:25-skyrmions}b shows a subset of the breathing modes in the
20\,GHz band: the three lowest frequency modes, the mode in the centre of the
band, and the four highest frequency modes. The lowest mode at 17.8\,GHz shows
alternating out-of-phase excitation of neighbouring skyrmions, at the highest
mode at 21.08\,GHz all skyrmions breathe in-phase. At the centre mode
(19.49\,GHz) only every second skyrmion is excited and neighbouring excited
skyrmions are out of phase. The intensity modulation seen over the whole the
strip follows model~(\ref{eq:coupled-oscillator-amplitude}).

To demonstrate signal propagation along the SAF skyrmion chain we now switch to
a different type of excitation and focus on the top bottom in-phase breathing
band around 20\,GHz. Instead of a sinc-pulse with equal power for all
frequencies up to 50\,GHz, we excite the left-most skyrmion with a Gaussian
pulse centred at $f_0=20.5$\,GHz:
\begin{equation}
  \mathbf{H}(t) = H_0 \sin(2 \pi f_0 t) \exp\left(-\frac{(t-t_0)^2}{2 \sigma_\mathrm{f}^2} \right)\mathbf{\hat{z}},
\end{equation}
where $t_0=0.4$\,ns is time of maximum excitation, $\sigma_\mathrm{f}=0.1$\,ns
controls the width of the excitation in frequency space and $\mu_0H_0=1$\,mT is the
maximum excitation strength. We use damping $\alpha=10^{-4}$. The central frequency
$f_0 = 20.5$\,GHz is chosen somewhat arbitrarily and not central in the band.
The exact location is irrelevant provided the overall pulse overlaps with the
breathing band. The dashed line in Fig.~\ref{fig:25-skyrmions}c shows the pulse
profile in frequency space (scaled by an arbitrary factor; the unusual shape of
the gaussian profile is caused by the logarithmic y-axis). We evolve the LLG for
1\,ns with applied excitation, i.e. $\pm4\sigma_\mathrm{f}$ plus an extra
0.2\,ns at the end of the excitation. The excitation strength at the beginning
and end of this time frame is very close to zero. After the excitation we record
data of the free LLG relaxation for further 20\,ns.

Figure~\ref{fig:25-skyrmions}c shows a power spectrum for the free relaxation,
i.e. the last 20\,ns. The whole top bottom in-phase breathing band has been
excited. Furthermore, we can observe weak excitation of parts of the
low-frequency translational modes. The higher-frequency top bottom out-of-phase
breathing band is not excited.

Figure~\ref{fig:25-skyrmions}d shows the signal propagation through the strip
during the first 7\,ns (including the excitation). We plot the deviation $\Delta
m_\mathrm{z}$ from the equilibrium configuration in the bottom layer averaged
over the strip width. The vertical stripe-like features correspond to the 25
skyrmions. We can see the strong excitation of the left-most skyrmion during the
first $\sim 0.8$\,ns. After that the signal propagates through the strip (to the
right in the figure) until it reaches the right strip edge at around 3\,ns and
is reflected. Shortly before 6\,ns the reflected signal arrives back at the left
strip edge. We can see that the excited modes seem to have slightly different
velocities and the excitation pulse disperses over time (sub-dominant features
in the figure after the first excitation). Reflection at the sample edges seems
to contribute significantly to the peak broadening.

Figure~\ref{fig:25-skyrmions}e shows the excitation of skyrmions~1 (left strip
edge, excited by the pulse), 13 (strip centre), and 25 (right strip edge) in
more detail. We can see the high-frequency oscillation of the carrier frequency
of the excitation pulse. The low-frequency envelope corresponds to the signal
propagating through the strip. By tracking the first maximum, we can compute the
average speed of the signal, and obtain approximately $300\,\mathrm{m/s}$.
This
is about 15\% faster than for a ferromagnetic system of similar geometry, for
which we obtain approx. $255\,\mathrm{m/s}$ (Supplementary Figs.~10 and~11).
Comparable velocities for ferromagnetic skyrmions have previously been reported in literature~\cite{kim2018a}.

\section{Discussion}\label{sec:summary}

We have systematically investigated the normal modes of confined
synthetic-antiferromagnetic (SAF) skyrmions and SAF skyrmion chains using
micromagnetic eigenvalue and ringdown simulations.

For a single skyrmion in a confined ferromagnetic (FM) geometry, we recover the
characteristic hierarchy of eigenmodes previously reported for confined chiral
skyrmions: a low-frequency gyrotropic mode, an intermediate breathing mode, and
a higher-frequency rotational mode with opposite handedness compared to the
gyrotropic mode. In square geometries, the gyrotropic trajectories remain
approximately circular, whereas rectangular confinement lifts the rotational
symmetry through direction-dependent restoring forces and leads to elliptical
trajectories.

The SAF geometry fundamentally changes the character of the gyrotropic
excitations. In contrast to the FM case, where one rotational direction is
energetically preferred for a given skyrmion polarity, the SAF consists of two
antiferromagnetically coupled skyrmions with opposite core polarities. As a
result, the preferred gyration direction of one layer competes with that of the
other in this frustrated dynamical system. In a square geometry, this
frustration manifests itself in the appearance of two closely spaced gyrotropic
eigenmodes with lifted degeneracy. These two modes are the lowest frequency
modes in the SAF skyrmion system at around $\approx 4.7\,$GHz, whereas the
lowest frequency mode in the ferromagnetic system is at $\approx 0.85\,$GHz.

We reveal that modifying the confinement from a square to a rectangle introduces
another qualitative transition: the gyration motion, where top and bottom layers
exhibit the same clockwise rotation in the square geometry, evolves into
clockwise rotation in one layer and counterclockwise rotation in the other. The
opposite rotation senses combine to produce translational modes. These
translational modes appear as the low-frequency excitations in arrays of
skyrmion, where the sample geometry is likewise anisotropic.

The breathing dynamics are comparatively less affected by the SAF coupling. For
both two-skyrmion and five-skyrmion systems, we identify breathing modes
qualitatively similar to those reported for FM systems. However, the SAF
geometry additionally supports out-of-phase breathing modes between the top and
bottom layer, reflecting the additional degree of freedom introduced by the
antiferromagnetic bilayer structure. These modes have no direct analogue in
single-layer FM skyrmions.

For chains containing many skyrmions, the collective modes exhibit
standing-wave-like spatial profiles. Translational horizontal and breathing
modes are well described by a standing-wave model, but the model cannot describe
the translational vertical modes.

Finally, we demonstrate signal propagation along extended SAF skyrmion chains.
The observed propagation velocities are comparable to those reported for
ferromagnetic skyrmion chains, despite the fundamentally different internal
dynamics of the SAF skyrmions. This suggests that SAF skyrmion chains may
support information transport with similar performance while simultaneously
benefiting from reduced stray fields and suppression of the skyrmion Hall
effect.

Synthetic-antiferromagnetic skyrmion chains constitute a rich and controllable
system for studying collective spin dynamics and skyrmion-based microwave,
magnonic and neuromorphic functionalities.

\section{Methods} \label{sec:method}

\subsection{Micromagnetic model}

We study the dynamic properties of a SAF skyrmion chain using the micromagnetic
model.
We represent the magnetisation vector field $\mathbf{M}(\mathbf{r}, t)$ as
$\mathbf{M}(\mathbf{r}, t) = M_\mathrm{s} \mathbf{m}(\mathbf{r}, t)$ where
$M_\mathrm{s}$ is the spontaneous magnetisation and $\mathbf{m}(\mathbf{r}, t)$
is the normalised magnetisation vector field that depends on space $\mathbf{r}$
and time $t$ with $|\mathbf{m}(\mathbf{r}, t)|=1$. The total energy $E$ is given
as:
\begin{equation}
  E = \int \left( w_{\mathrm{ex}} + w^{z}_{\mathrm{dmi}} +
  w_{\mathrm{dem}} + w_{\mathrm{K}} + w_{\mathrm{RKKY}}\right)\, \mathrm{d}^3 r,
\end{equation}
where energy density \(w_\mathrm{ex}\) is the isotropic exchange contribution,
\( w_\mathrm{dmi}^z\) is the interfacial Dzyaloshinskii-Moriya interaction (DMI)
contribution, \(w_\mathrm{dem}\) is the magnetostatic field contribution,
\(w_\mathrm{K}\) is the perpendicular uniaxial anisotropy contribution, and
\(w_{\mathrm{RKKY}}\) is the RKKY interaction contribution. All energy density
terms depend on the magnetisation~$\mathbf{M}$~\cite{bruckner2023}.

The dynamics is described by the Landau-Lifshitz-Gilbert (LLG) equation
\begin{equation}
  \label{eq:llg}
  \frac{\mathrm{d}\mathbf{m}}{\mathrm{d}t}
  =
  -\gamma' \left(
    \mathbf{m} \times \mathbf{H}_{\mathrm{eff}}
    + \alpha \, \mathbf{m} \times \left( \mathbf{m} \times
    \mathbf{H}_{\mathrm{eff}} \right)
  \right),
\end{equation}
where $\gamma' = \frac{\gamma}{1+\alpha^2}$ and $\gamma$ denotes the
gyromagnetic constant, $\alpha$ is the Gilbert damping parameter, and
$\mathbf{H}_{\mathrm{eff}}$ is the effective magnetic field arising from the
variational derivative of the total energy with respect to the magnetisation
\begin{equation}
  \label{eq:heff}
  \mathbf{H}_{\mathrm{eff}}(\mathbf{m})
  =
  - \frac{1}{\mu_0 M_s}\,
  \frac{\delta E[\mathbf{m}]}{\delta \mathbf{m}} .
\end{equation}

\subsection{System geometry and material parameters}\label{sec:methods-geometry}

Figure~\ref{fig:geometries} illustrates the geometries studied in this work. All
systems are rectangular thin films. In the ferromagnetic case the film has
thickness 2\,nm. We model the SAF as two ferromagnetic layers, referred to as
top and bottom layer, which are anti-ferromagneticaly coupled via the RKKY
interaction~\cite{ruderman1954}. We do not explicitly model the geometry of a
spacer layer. In most simulations the top and the bottom layer have a thickness
of 1\,nm each, and the total film thickness remains 2\,nm.

The material parameters follow~\cite{lonsky2020a}, where a single skyrmion was
studied in a SAF material with cylindrical geometry: exchange stiffness $A =
15$\,pJ/m, perpendicular uniaxial anisotropy along~z with $K_\mathrm{u} =
1$\,MJ/m\(^3\), RKKY coupling constant $\sigma = -3 \times
10^{-4}\,\text{J/m}^2$, interfacial DMI coupling constant $D =3\,\text{mJ/m}^2$,
and spontaneous magnetisation $M_\mathrm{s} = 0.795\,\text{MA/m}$ (corresponding
to $\mu_0M_\mathrm{s} = 1\,\text{T}$).

\subsection{Implementation of normal mode computation}

To numerically solve the micromagnetic model, we use
\texttt{magnum.np}~\cite{bruckner2023}, an open-source, Python-based
finite-difference micromagnetic solver built on the PyTorch
framework~\cite{pytorch2019}, which can run on either CPU or GPU. We use cubic
discretisation cells of (\(1 \times 1 \times 1\))\,nm$^3$.

We compute normal modes in this work using two methods: the ringdown method,
which carries out Fourier analysis in the time domain~\cite{mcmichael2005}, and
an eigenvalue method, which linearises the LLG before solving a generalised
eigenvalue problem~\cite{daquino2009}. The simulation software
\texttt{magnum.np} supports the necessary time integration and computation of
the eigenvectors for both methods.

Both methods require a relaxed magnetisation configuration corresponding to a
local energy minimum. We initialise the magnetisation in a state comprising of
Néel-type skyrmions with opposite polarity in each layer. To relax the system
into a local energy minimum, we carry out a time integration of the LLG equation
for 10\,ns with damping $\alpha = 0.1$. The somewhat high damping extracts
energy from the system to assist finding an energy minimum. The resulting state
is the equilibrium state from which we continue with either the ringdown or the
eigenvalue method. We denote the equilibrium state with $\mathbf{m}_0$.

\subsubsection{Ringdown method}\label{sec:ringdown-method}

Starting from the equilibrium state, the ringdown method has two consecutive
phase: 1. the excitation of the system with a small-amplitude applied magnetic
field, and 2. the recording of the `ringing-down' of the system, which is used
for the time-domain Fourier analysis. In both phases, the damping is set to a
much smaller value of $\alpha=10^{-3}$ to avoid suppression of the dynamics.

The excitation in the ringdown method is chosen to be an external magnetic field
that is (in most of our simulations) pointing in the z-direction (perpendicular
to the SAF stack) and a sinc pulse in the time domain:
\begin{equation}
\mathbf{H}_\mathrm{z} (t) = H_0\,\text{sinc}(2 \pi
f_\mathrm{c}(t-t_\mathrm{p})) \mathbf{\hat{z}}
\end{equation}
with cutoff frequency $f_\mathrm{c}=50\,\text{GHz}$. The peak of the sinc pulse
is reached after the amount of time $t_\mathrm{p} = 1\,\text{ns}$. The total
simulated time for the excitation phase is 5$\,$ns. We choose a spatial
excitation profile such that we excite the full system when there is a single
(SAF) skyrmion. For multiple SAF skyrmions, we only excite the left top-bottom
pair in the strip. The excitation field is spatially uniform in the rectangular
cuboidal region containing the left SAF skymion, exciting both the skyrmion and the
background.

Following the excitation phase, the ringing-down phase starts, during which we
record data for subsequent Fourier analysis. This ringing-down phase is 20$\,$ns
long, and we record the magnetisation configuration in $5\,\text{ps}$ steps.

We analyse the data recorded during the relaxation phase using the spatially
resolved method~\cite{beg2017a}. We first subtract the static background
$\mathbf{m}_0$, to focus on the change $\Delta \mathbf{m}(t) = \mathbf{m}(t) -
\mathbf{m}_0$ of the magnetisation (and to eliminate a zero frequency peak in
the Fourier transform). We then apply a discrete Fourier transform (DFT) to
\(\Delta \mathbf{m}\) over time:
\begin{equation}
\mathcal{F}_k(\mathbf{r}_i, f) = \sum_{j=1}^{n} \Delta
m_k(\mathbf{r}_i, t_j) \exp(-2\pi i f t_j), \label{eq:fourier-amplitude}
\end{equation}
where $k \in \{x, y, z\}$ denotes the vector component and $n=4000$ is the
number of time steps. We then compute the power spectral density (PSD):
\begin{equation}
  \label{psd_eq}
  \mathrm{PSD}_k(f) = \frac{1}{N} \sum_{i=1}^{N} |\mathcal{F}_k(\mathbf{r}_i, f)|^2,
\end{equation}
where $N$ is the total number of cells in the finite difference grid.

\subsubsection{Eigenvalue method}\label{sec:eigenvalue-method}

For the eigenvalue method we need to compute eigenvalues and eigenvectors of the
linearised LLG equation for small perturbations of the equilibrium configuration
$\mathbf{m}_0$~\cite{daquino2009}. For this we use the eigenvalue solver built
into \texttt{magnum.np}, which internally uses the ARPACK implementation of the
implicitly restarted Arnoldi method~\cite{MaschhoSorensen2005} via the SciPy
library~\cite{virtanen2020}. We obtain both eigenvalues and complex
eigenvectors.

From the eigenvectors we can compute the PSD following~\cite{daquino2009,
  daquino2023}. This can be done in two different ways: 1.~The PSD can be
computed from the equilibrium configuration and a given spatial profile of the
excitation field. This method is computationally most efficient. 2. We can
calculate a deviated equilibrium configuration including the excitation field.
The PSD is then computed from the difference of the deviated and the original
equilibrium configuration. This method is computationally more expensive. We
have implemented both methods in \texttt{magnum.np}. For more details refer to
Supplementary Sec.~V. Both methods produce PSDs
with peaks at identical frequencies, but the relative power of the modes can
vary between the two methods. We use method 2 throughout this work, as its
resulting power distribution is closer to a full ringdown simulation.

\subsection{Data analysis}\label{sec:data-analysis}

To analyse the dynamics of the mode associated with frequency $f$, we can
compute the dynamics of the magnetisation vector at position $\mathbf{r}$ using:
\begin{equation}
  m_k(\mathbf{r}, t, f) = m_{0,k}(\mathbf{r})
  + B\,\Re[\mathcal{F}_k(\mathbf{r}, f) \exp(2\pi i f t)],
  \label{eq:mode_reconstruction}
\end{equation}
where $B$ is an artificial scaling factor to enhance the visibility of the
eigenmode in real-space visualisations. We use $B = 0.5 /
\max(\Re[\mathcal{F}_k(\mathbf{r}, f) \exp(2\pi i f t)])$ throughout this work.

To visualise the motion of the skyrmions for a given eigenmode of the system, we
compute the first moment of the topological charge density to track the
skyrmion's core per z~layer. From the reconstructed magnetisation vector
field~(\ref{eq:mode_reconstruction}) we can compute the local topological charge
density:
\begin{equation}
q(\mathbf{r}, t)
=
\frac{1}{4\pi}
\mathbf{m}
\cdot
\left(
\frac{\partial \mathbf{m}}{\partial x}
\times
\frac{\partial \mathbf{m}}{\partial y}
\right),
\label{eq:topological_density}
\end{equation}
where the spatial derivatives are evaluated numerically using central
differences on the interior of the simulation grid. Similar
to~\cite{schutte2014} we track the skyrmion's core position $(R_\mathrm{x},
R_\mathrm{y})$ through the first moment of the topological charge density:
\begin{equation}
R_k(t) = \frac{\sum_i k_i \, q(\mathbf{r}_i, t)}{\sum_i q(\mathbf{r}_i, t)},
\label{eq:skyrmion_position}
\end{equation}
where the summation runs over all $\mathbf{r}_i = (x_i, y_i, z)$ in the layer
and $k \in \{x, y\}$. For systems containing multiple skyrmions, the
computational domain is partitioned into spatial subregions, and
Eq.~\eqref{eq:skyrmion_position} is applied independently within each region to
extract individual trajectories.

The displacement from the equilibrium position $R_{k,\mathrm{eq}}$ is very small
and by just plotting $R_{k}(t)$ trajectories would barely be visible. Therefore,
for visualisation we scale them with a factor $\lambda$:
\begin{equation}
R_{k,\mathrm{vis}}(t) = R_{k,\mathrm{eq}} + \lambda \left(R_{k}(t) - R_{k,\mathrm{eq}}\right),
\end{equation}
where we need to take into account that the skyrmion in the equilibrium
configuration is generally not located at the origin.

In SAFs, the top and bottom magnetic layers are aligned antiparallel at
equilibrium. This antiparallel alignment leads to a non-trivial interpretation
of phase data in the dynamic mode analysis: physically in-phase motion---where
both layers oscillate coherently relative to their own equilibrium
position---can appear as out-of-phase in the raw Fourier transformed data. To
ensure that in-phase dynamics are visually represented as such, we apply a
global phase shift of $\pi$ to the top layer before plotting. This transforms
the top layer’s phase as \(\phi^{'}_{\mathrm{top}} = \phi_{\mathrm{top}} +
\pi\), with values wrapped back into the [-$\pi$, $\pi$] interval.

\bibliography{SAF, introduction, arpack}

\section*{Acknowledgements} \label{sec:acknowledgements}

We thank Guido Meier for fruitful discussion.
This work is supported by the European Union’s Horizon research and innovation
program under MaMMoS No. 101135546, Marie Skłodowska Curie grant agreement No.
101152613, Higher Education Commission and Deutscher Akademischer
Austauschdienst grant No. 57630247 and in part by the Austrian Science Fund
(FWF) [10.55776/PAT3864023].

\section*{Author contributions}

KZ, SH, ML and HF conceived the study. FB implemented PSD calculation in
\texttt{magnum.np} and contributed Supplementary Sec.~V. KZ and ML performed the simulations and analysed the data.
HF, ML and KZ wrote the manuscript. All authors contributed to discussing the
results.

\section*{Data availability}

Data and code to reproduce all results will be made available together with the
final publication.

\section*{Competing interests}

The authors declare no competing interests.

\end{document}

\title{Eigenmodes of synthetic antiferromagnetic skyrmions -- Supplementary material}
\date{\today}

\author{Kauser Zulfiqar}
\affiliation{Max Planck Institute for the Structure and Dynamics of Matter, Hamburg, Germany.}
\affiliation{Center for Free-Electron Laser Science, Hamburg, Germany.}
\affiliation{Department of Physics, University of Hamburg,
  Hamburg, Germany.}

\author{Martin Lang}
\affiliation{Max Planck Institute for the Structure and Dynamics of Matter, Hamburg, Germany.}
\affiliation{Center for Free-Electron Laser Science, Hamburg, Germany.}

\author{Samuel Holt}
\affiliation{Max Planck Institute for the Structure and Dynamics of Matter, Hamburg, Germany.}
\affiliation{Center for Free-Electron Laser Science, Hamburg, Germany.}

\author{Swapneel Amit Pathak}
\affiliation{Max Planck Institute for the Structure and Dynamics of Matter, Hamburg, Germany.}
\affiliation{Center for Free-Electron Laser Science, Hamburg, Germany.}

\author{Florian Bruckner}
\affiliation{University of Vienna, Vienna, Austria}

\author{Hans Fangohr}
\affiliation{Max Planck Institute for the Structure and Dynamics of Matter, Hamburg, Germany.}
\affiliation{Center for Free-Electron Laser Science, Hamburg, Germany.}
\affiliation{University of Southampton, Southampton, UK}

\maketitle

\section{Selecting a suitable colormap for visualising phase amplitude maps}

We use phase amplitude maps to visualise the eigenmodes. In these visualisations
the phase of the complex eigenmodes is shown with the hue. The amplitude is shown
via the alpha channel, such that only high-intensity regions contribute to the
visualisation. For these visualisations to work without artifacts it is crucial
to select a colourmap for the hue that has uniform lightness. The commonly used
``hsv'' colormap does not fulfil this requirement and introduces artifacts. We
instead use the ``husl'' colormap from \href{https://seaborn.pydata.org/tutorial/color\_palettes.html}{seaborn}.

\begin{figure}[h]
  \includegraphics[width=\linewidth]{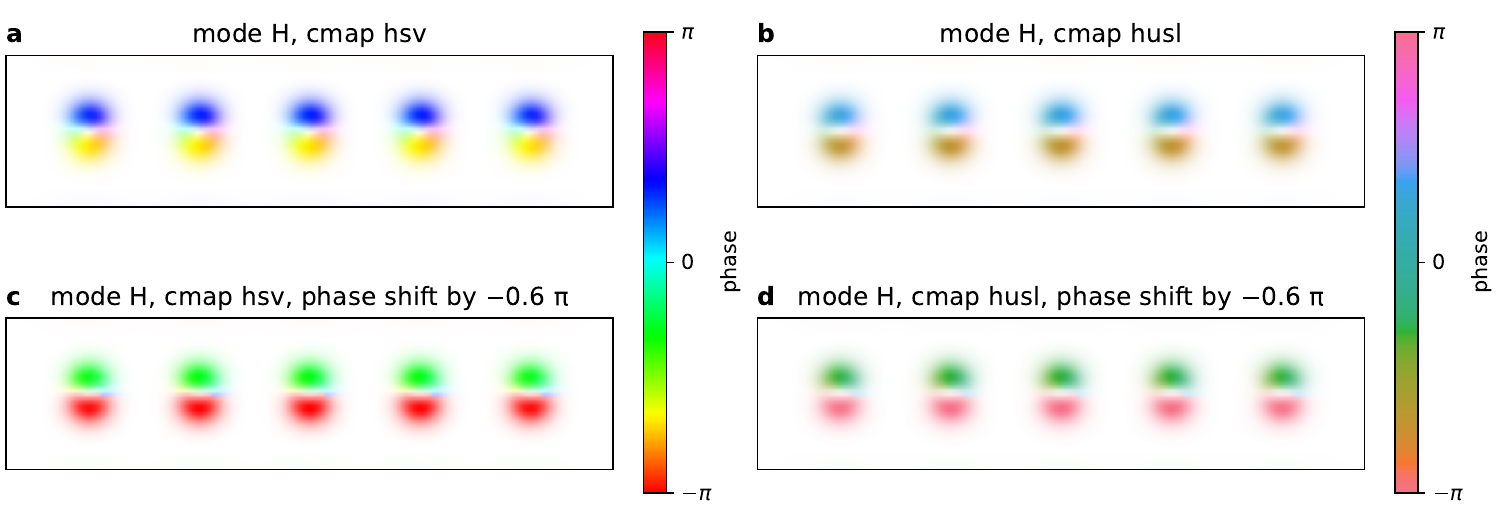}
  \caption{\label{fig:importance-of-lightness}Phase amplitude map for the same
    translational mode of five skyrmions with four different visualisations. (a) hsv to visualise the phase, (b)
    husl to visualise the phase, (c) hsv with all colours shifted by $-0.6\pi$,
    (d) husl with all colours shifted by $-0.6\pi$.}
\end{figure}

Figure~\ref{fig:importance-of-lightness} shows the effect for a selected
mode: vertical translation of five skyrmions, taken from Fig.~7 of the main text. The same mode is displayed four
times, twice with hsv colourmap (a and c), and twice with husl colourmap (b and
d). For the subfigures c and d all colours have been shifted by $-0.6\pi$.
(The overall phase has no meaning, only phase differences are relevant.)
For the husl colourmap (subfigures~b and~d), the mode looks identical, we can
clearly see the translation in y direction and a strong suppression in x
direction. We can draw the same conclusion from subfigure~c, hsv with a well
chosen phase shift. We would however draw different conclusions from
subfigure~a, hsv without any phase shift. Here the mode looks much more like a
gyrotropic mode---which is misleading.

This example illustrates that homogeneous hue lightness is crucial to
draw valid conclusions from the phase amplitude plots. Using
colourmaps such as hsv that ignore this constraint can lead to
artifacts in the visualisation.

\clearpage
\section{Five skyrmions}

\begin{figure}[hbp]
	\includegraphics[width=\linewidth]{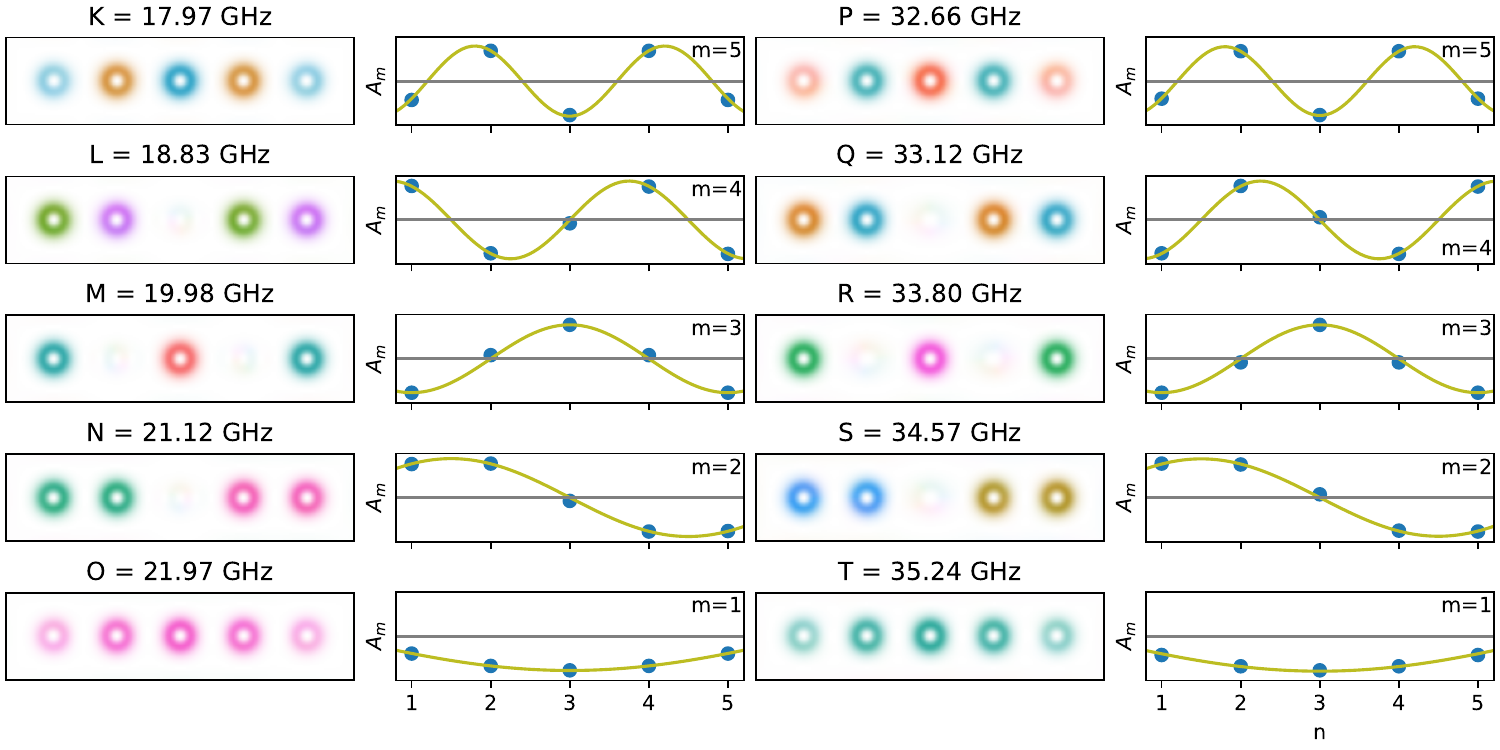}
  \caption{Breathing modes~K to~T of the five-skyrmion system discussed in
    Fig.~7 in the main text. The phase amplitude maps show the top layer. For
    modes~K to~O both layers breathe in phase, for modes~P to~T they breathe out
    of phase (see Fig.~\ref{fig:modes-five-skyrmions} for phase amplitude maps
    showing both layers). The plots in columns two and four show
    amplitudes $A_m(n)$ of collective skyrmion coordinates from
    simulation (blue points) and amplitudes $\tilde{A}_m(n)$
    (Eq.~1 in the main text) of the associated standing
    mode $m$ (green line).
  }
\end{figure}

\clearpage
\section{Eigenmodes of FM and SAF skyrmions}

The following Figs.~3 to~8 show phase amplitude maps for the first 40 eigenmodes of
FM and SAF skyrmions in different system geometries.

\begin{figure}[hbp]
  \includegraphics[width=0.95\linewidth]{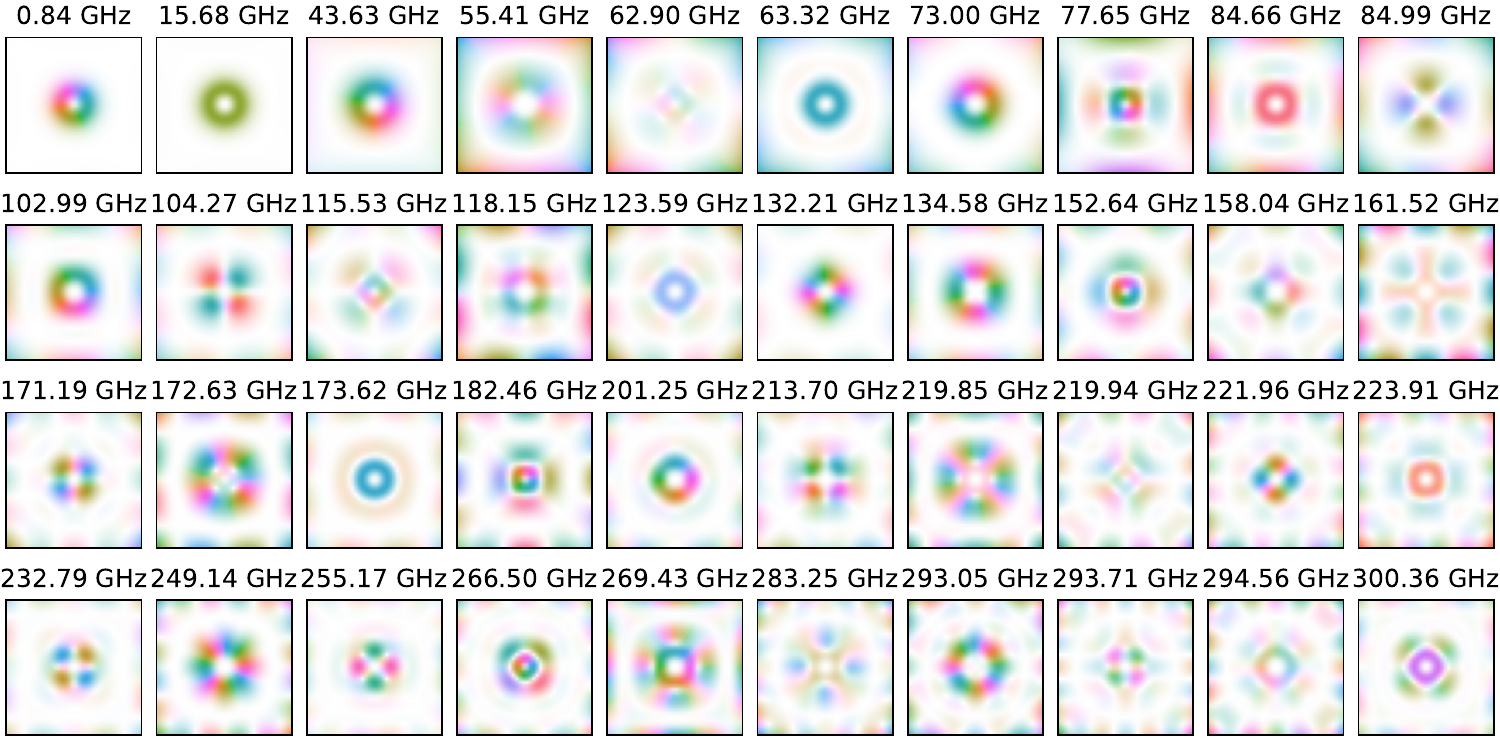}
  \caption{Eigenmodes of a single FM skyrmion in a $(40 \times 40 \times
    2)\,\mathrm{nm}^3$ geometry. The first three modes are discussed in detail
    in Fig.~2 in the main text.
  }
\end{figure}

\begin{figure}[hbp]
  \includegraphics[width=0.95\linewidth]{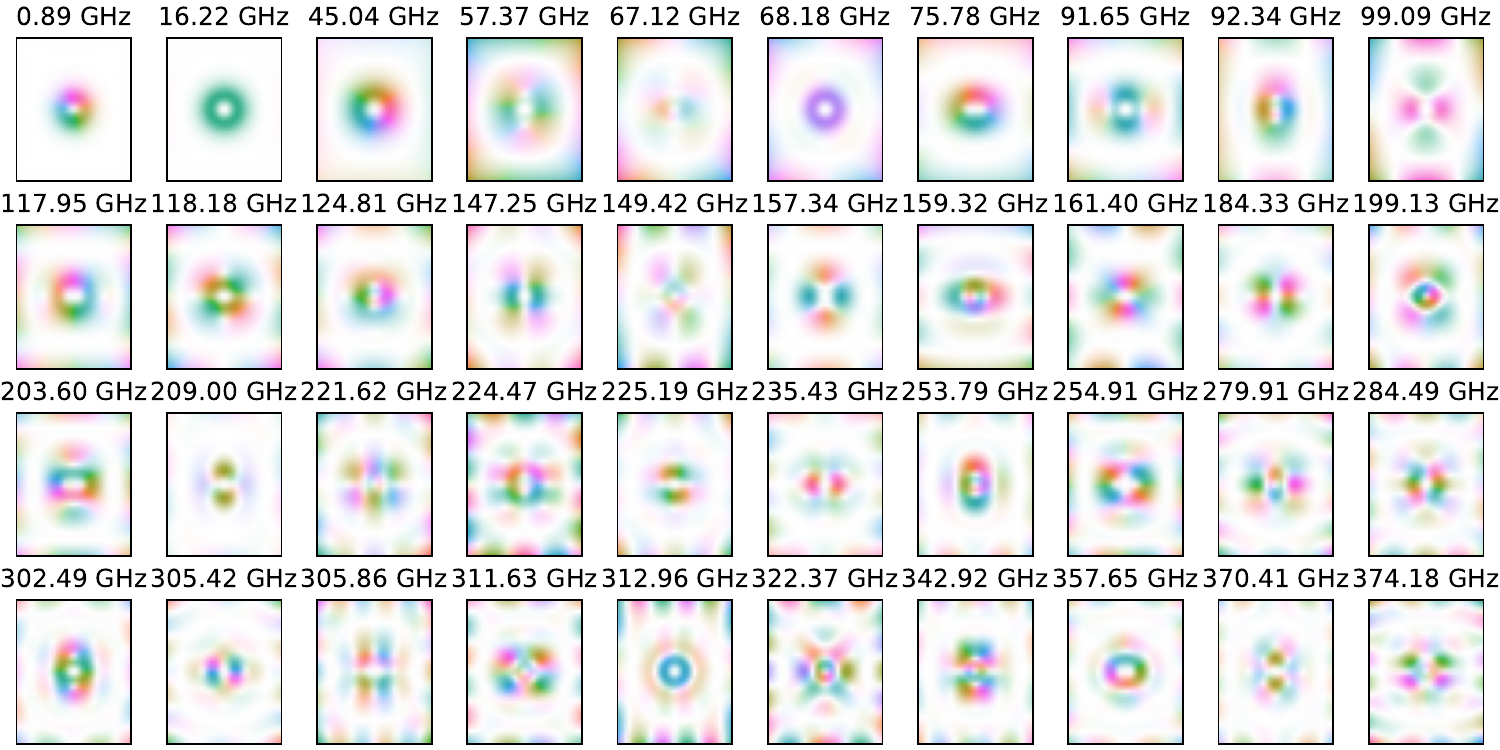}
  \caption{Eigenmodes of a single FM skyrmion in a $(32 \times 40 \times
    2)\,\mathrm{nm}^3$ geometry. Compared to the square geometry, many of the
    modes show elliptical distortion, see Fig.~\ref{fig:single-skyrmion-fm-square-rectangle-comparison} for
  a more detailed comparison of the gyrotropic and first rotational mode.}
\end{figure}

\begin{figure}
	\sffamily\textbf a

	\includegraphics[width=0.5\linewidth]{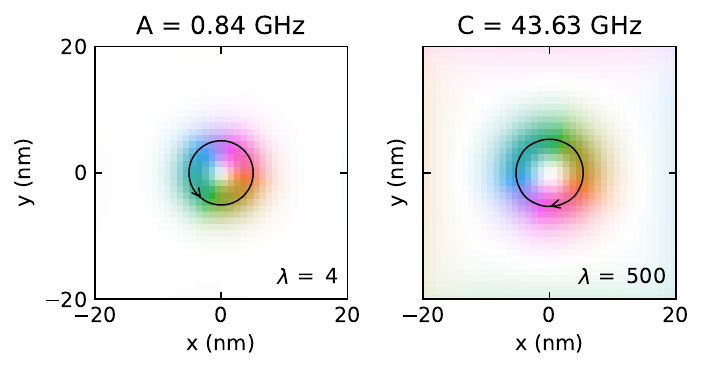}

	\textbf b

	\includegraphics[width=0.5\linewidth]{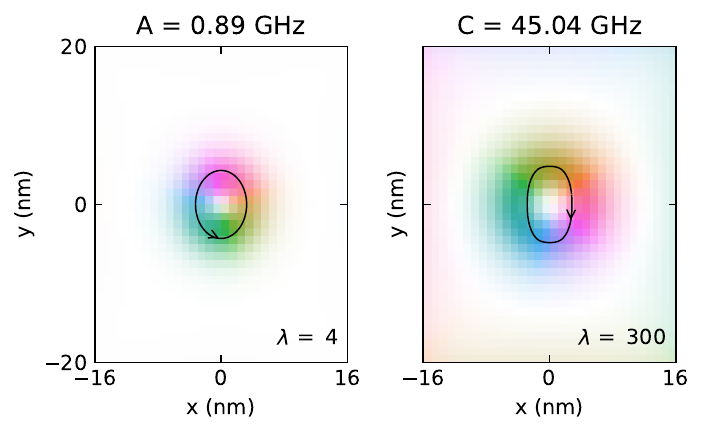}

  \caption{Comparison of the low-frequency gyrotropic and rotational modes for a FM skyrmion in
    (a) square geometry of size $(40 \times 40 \times 2)\,\mathrm{nm}^3$
    and (b) rectangular geometry of size $(32 \times 40 \times 2)\,\mathrm{nm}^3$.
    The trajectories show that in the square geometry the
    skyrmion moves in a circle, while in the rectangular geometry the motion is
    elliptical.
    \label{fig:single-skyrmion-fm-square-rectangle-comparison}
  }
\end{figure}

\begin{figure}
  \includegraphics[width=\linewidth]{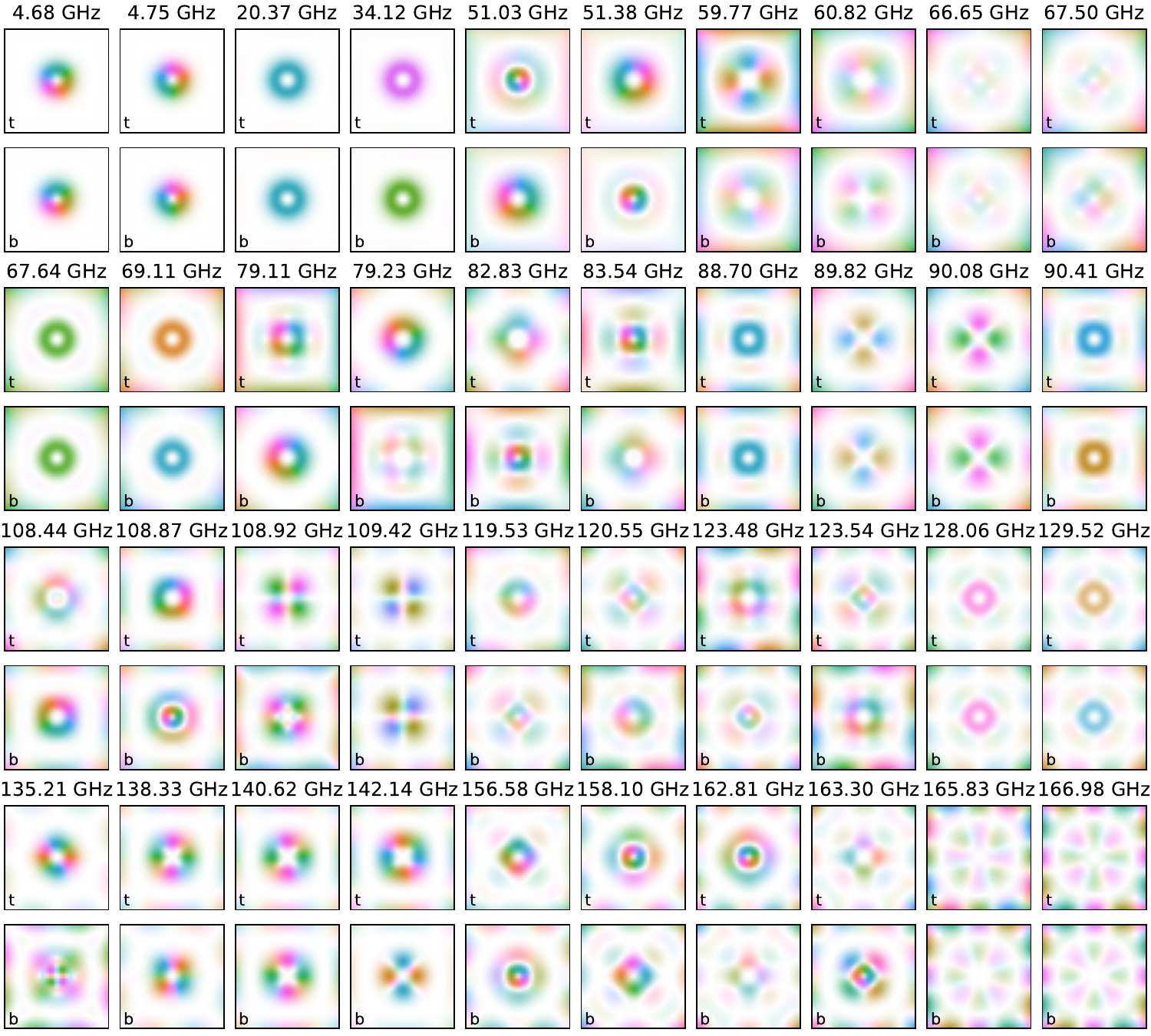}
  \caption{Eigenmodes of a single SAF skyrmion in a $(40 \times 40 \times
    2)\,\mathrm{nm}^3$ geometry. The two layers are marked with letters t (top)
    and b (bottom). The first four modes are discussed in detail in Fig.~3
    in the main text.}
\end{figure}

\begin{figure}
  \includegraphics[width=\linewidth]{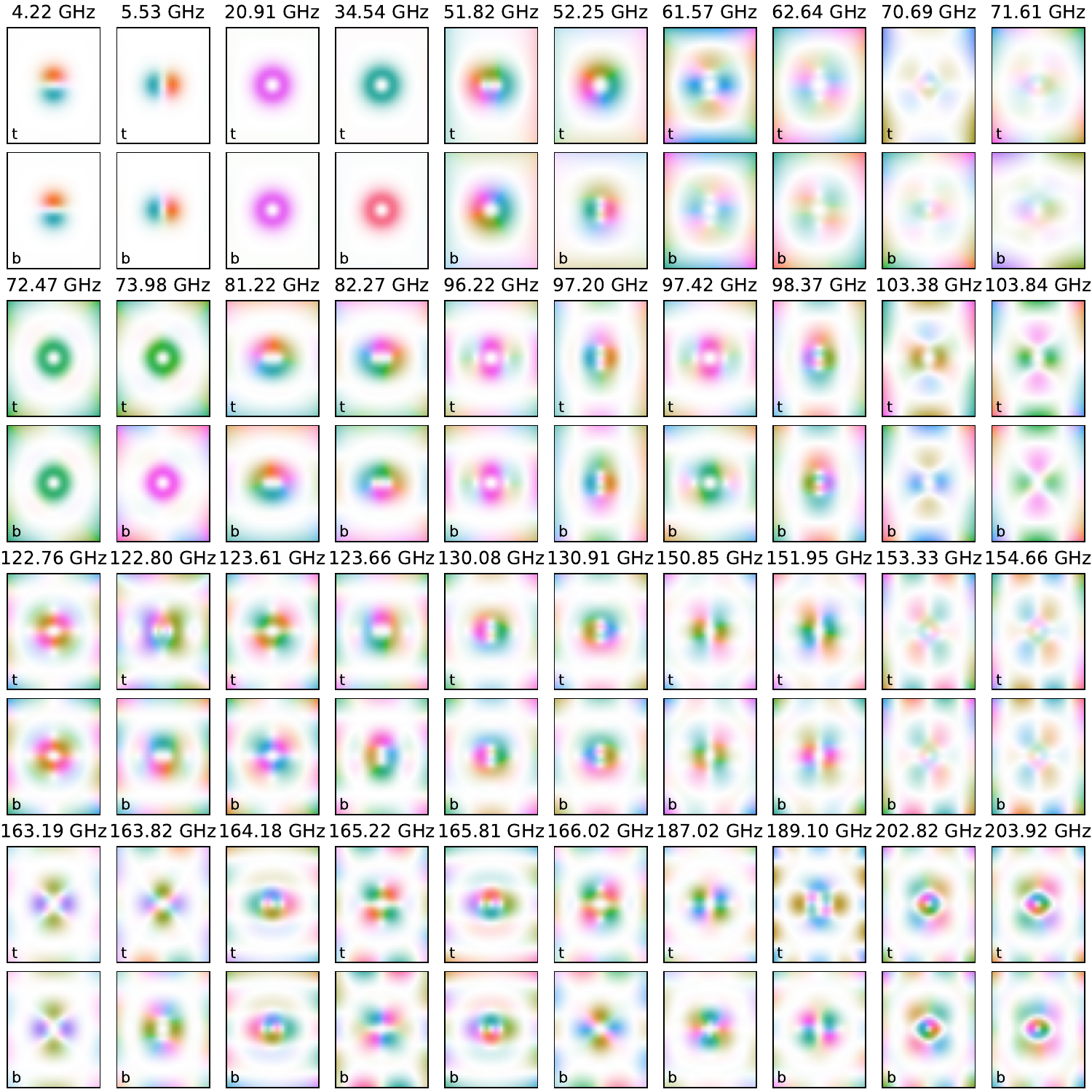}
  \caption{Eigenmodes of a single SAF skyrmion in a $(32 \times 40 \times
    2)\,\mathrm{nm}^3$ geometry. The two layers are marked with letters t (top)
    and b (bottom). The first four modes are discussed in detail in Fig.~4
    in the main text.}
\end{figure}

\begin{figure}
  \includegraphics[width=\linewidth]{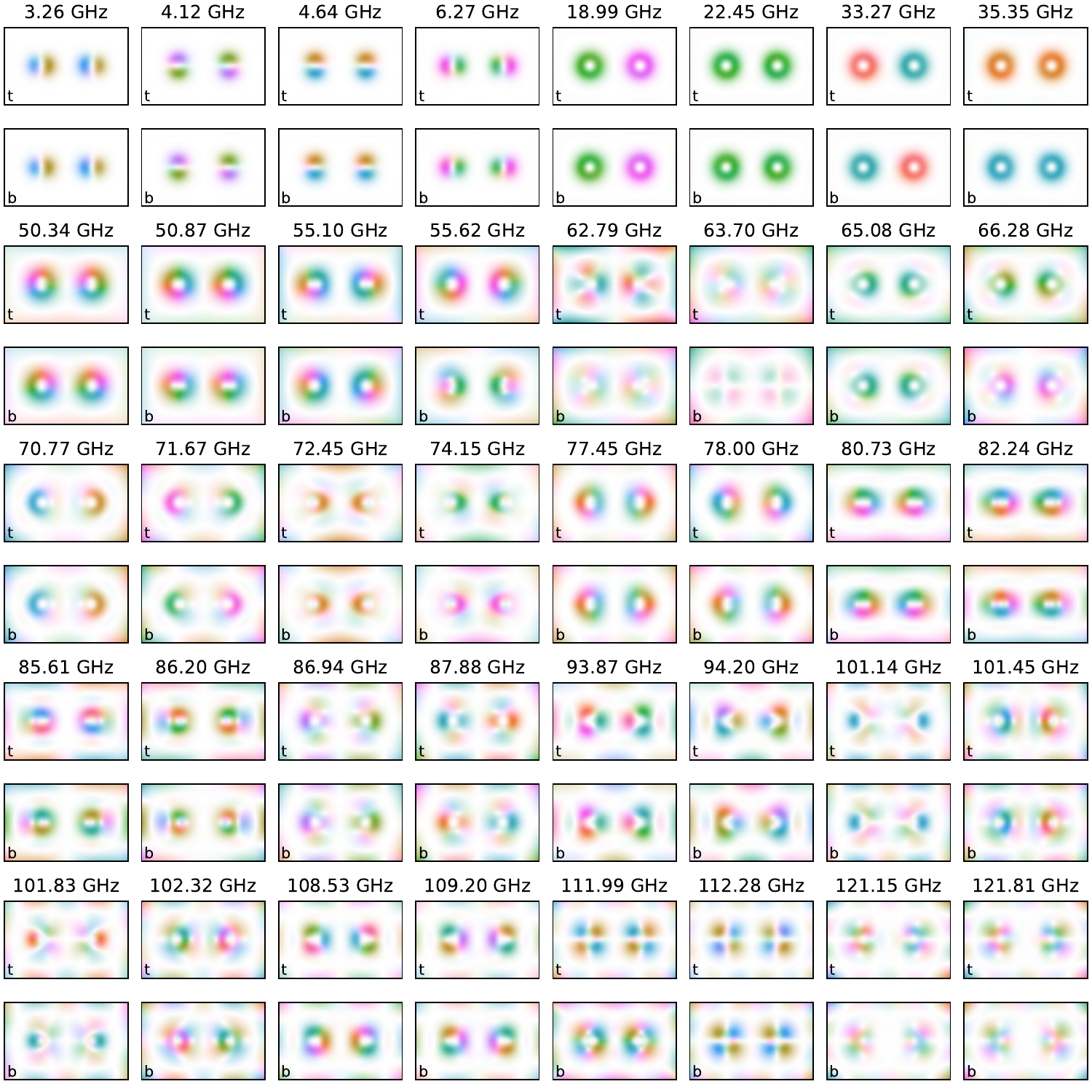}
  \caption{Eigenmodes of two SAF skyrmions in a $(64 \times 40 \times
    2)\,\mathrm{nm}^3$ geometry. The two layers are marked with letters t (top)
    and b (bottom). The first eight modes are discussed in detail in Fig.~6
    in the main text.}
\end{figure}

\begin{figure}
  \includegraphics[width=.95\linewidth]{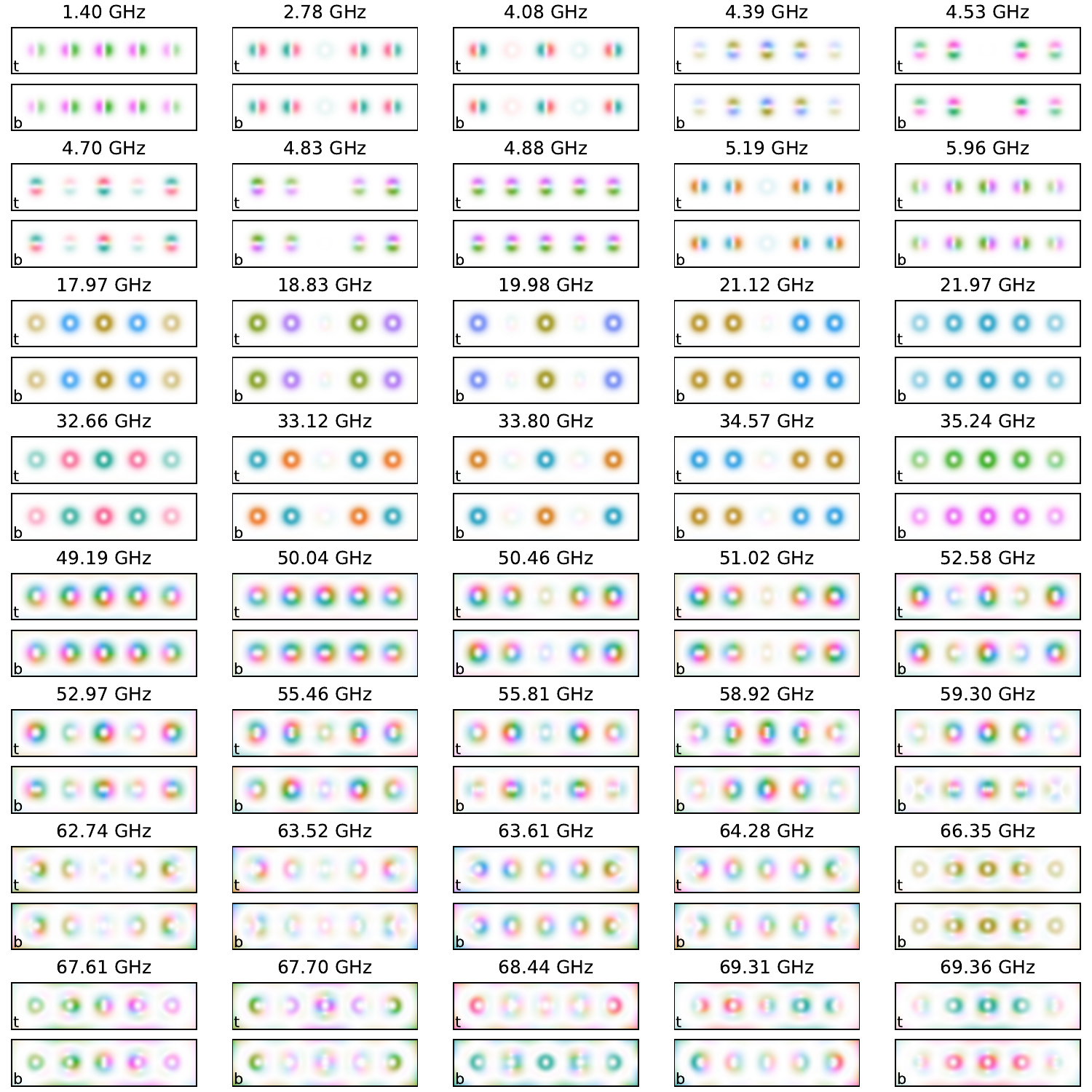}
  \caption{Eigenmodes of five SAF skyrmions in a $(160 \times 40 \times
    2)\,\mathrm{nm}^3$ geometry. The two layers are marked with letters t (top)
    and b (bottom). The first 20 modes are discussed in detail in Fig.~7
    in the main text.
    \label{fig:modes-five-skyrmions}
  }
\end{figure}

\clearpage
\section{Signal propagation in a FM strip}

\begin{figure}[hbp]
  \includegraphics[width=\linewidth]{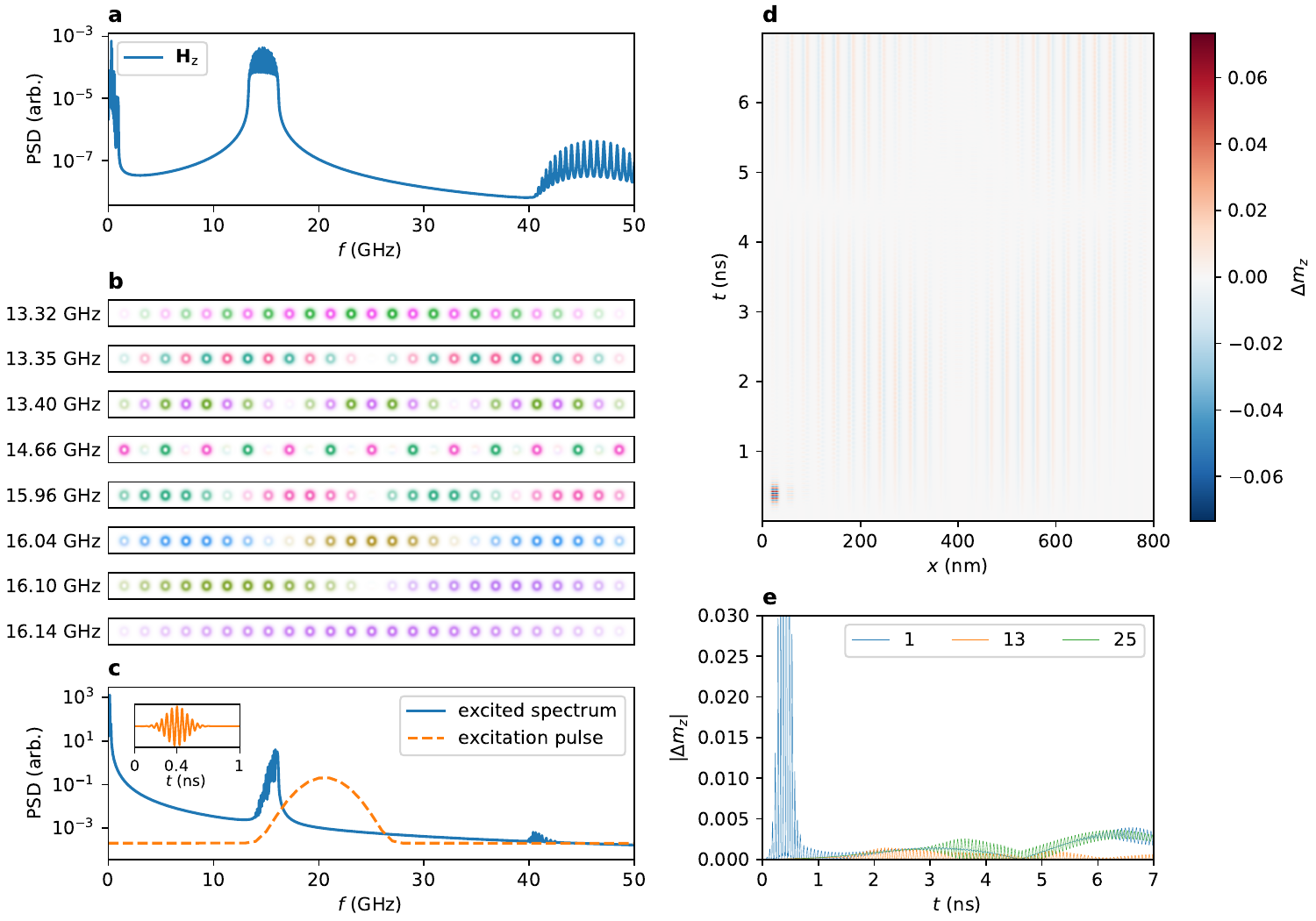}
  \caption{
    Signal propagation in a FM strip hosting 25 skyrmions, similar to Fig.~8 in
    the main text.
    (a) PSD for 25 skyrmions. (b) Subset of the breathing modes in the 15\,GHz
  band. Compared to the SAF system, this band has shifted to lower frequencies.
  (c) Spectrum after exciting the 15\,GHz band with the same gaussian pulse
  centered at 20.5\,GHz that has been applied to the SAF system in Fig.~8 in the
  main text. Only the left-most
  skyrmion is excited by the pulse. The pulse power distribution in frequency
  space is shown with the dashed line. The overlap of pulse and breathing band
  is narrow and the band is only excited relatively weakly and
  inhomogeneously (power decreases strongly as we go to lower frequencies in the
  band). Excitation with an adjusted pulse is shown in
  Fig.~\ref{fig:fm-excitation-15GHz}.
  Inset: shape of the excitation pulse in
  time domain.
  (d) Propagation of excitation through the strip. We do not see a strong signal
  propagating through the strip. Instead we can primarily observe the excitation
  of the first skyrmion at 0.4\,ns and weak excitation of most of the
  skyrmions with low-frequency modulation corresponding to the near zero frequency
  peak that dominates the spectrum show in subfigure~c. (e) Excitation of selected skyrmions in
  the strip (1: left-most skyrmion, 13: centre skyrmion, 25: right-most
  skyrmion). The applied excitation has its maximum at $t=0.4$\,ns.
  \label{fig:fm-excitation-20GHz}
}
\end{figure}

\begin{figure}[h]
  \includegraphics[width=\linewidth]{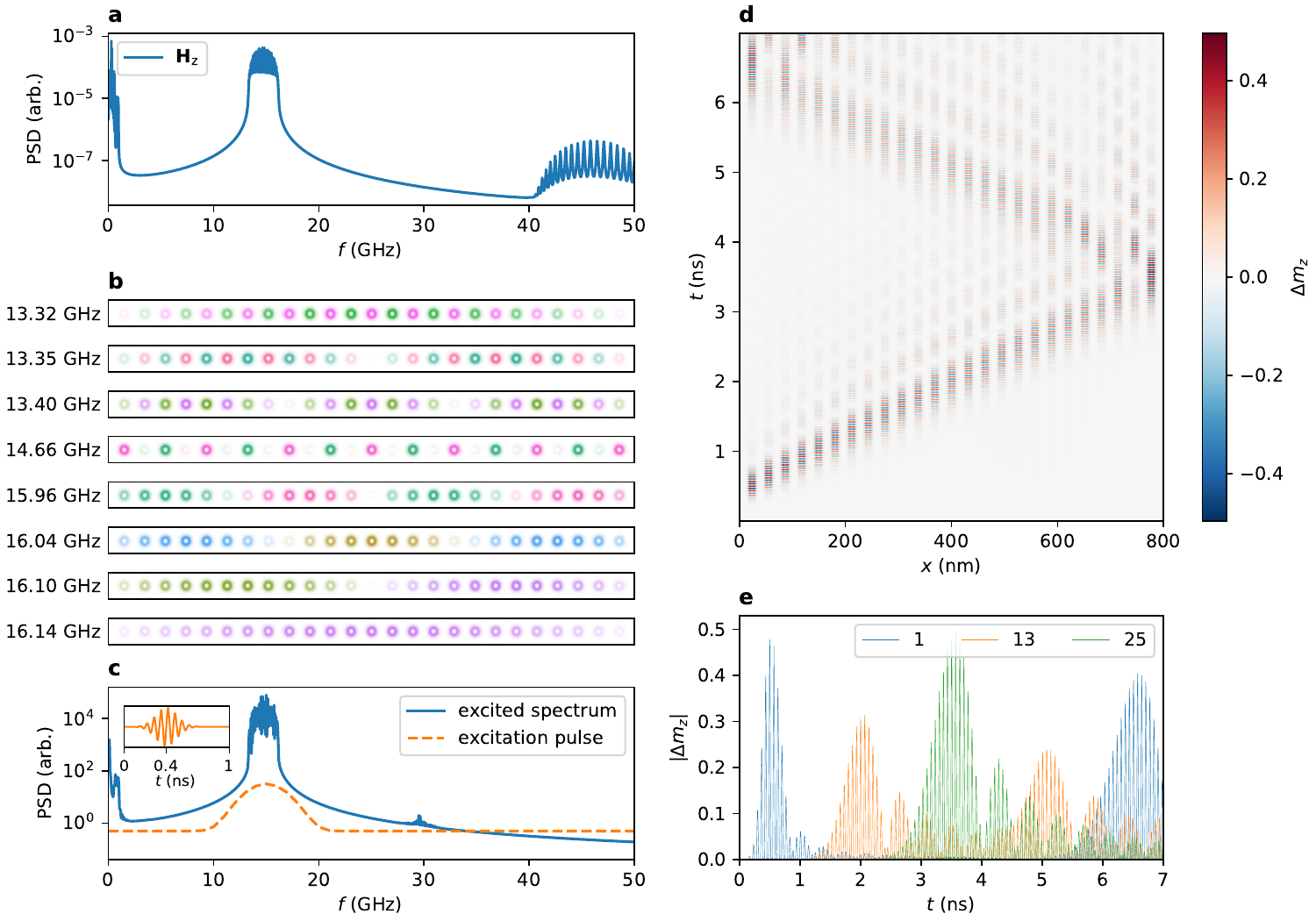}
  \caption{
    Signal propagation in a FM strip hosting 25 skyrmions, similar to Fig.~8 in
    the main text.
    (a) and (b) as in Fig.~\ref{fig:fm-excitation-20GHz} (c) Spectrum after
    exciting the 15\,GHz band with a gaussian pulse centred at 15\,GHz. Only the left-most
    skyrmion is excited by the pulse. The pulse power distribution in frequency
    space is shown with the dashed line. Inset: shape of the excitation pulse in
    time domain. Compared to Fig.~\ref{fig:fm-excitation-20GHz} the pulse
    overlaps properly with the breathing band and excitation of these modes
    dominates the spectrum.
    (d) Propagation of excitation through the strip. (e) Excitation of selected skyrmions in
    the strip (1:~left-most skyrmion, 13: centre skyrmion, 25: right-most
    skyrmion). The applied excitation has its maximum at $t=0.4$\,ns.
    By tracking the first maxima, we obtain a velocity of approx.~255\,m/s.
    \label{fig:fm-excitation-15GHz}
  }
\end{figure}

\clearpage
\section{Eigenmode solver in \texttt{magnum.np}}
\subsection{Introduction}
The dynamic response of magnetic nanostructures is governed by the Landau--Lifshitz--Gilbert (LLG) equation. A traditional approach to characterize the high-frequency properties of a magnetic system is to perform time-domain micromagnetic simulations driven by stochastic thermal fields~\cite{scholz_micromagnetic_2001, bruckner_large_2019} or pulsed excitations~\cite{baker_proposal_2017}, and to extract the resonance spectrum via Fourier analysis of the resulting magnetization dynamics~\cite{carlotti_exchange_2014}. While conceptually straightforward, this approach is computationally expensive, often requiring simulation times of several nanoseconds with sub-picosecond time steps, and yields noisy spectra that necessitate additional post-processing.

Theoretical approaches for the determination of small magnetization oscillations trace back to analytical treatments of saturated magnetic particles~\cite{bertram_thermal_2002, jin_quasi-analytical_2004}. For systems with arbitrary geometry and spatially inhomogeneous magnetization, numerical methods are required. The dynamical matrix method~\cite{grimsditch_magnetic_2004, mcmichael_magnetic_2005} computes all normal modes by diagonalizing a dense matrix, but its $\mathcal{O}(N^2)$ memory and computational cost limits it to small systems. An alternative matrix-free formulation was introduced by d'Aquino et al.~\cite{daquino_computation_2008, daquino_novel_2009, forestiere_finite_2009}, which recasts the linearized LLG equation as a generalized eigenvalue problem for self-adjoint operators. This formulation can be solved with iterative Krylov-subspace methods such as the Lanczos algorithm~\cite{lehoucq_arpack_1998}, requiring only matrix-vector products that leverage the same acceleration techniques used in time-domain micromagnetic codes. The approach has been extended to include semi-analytical perturbation techniques for damping and rf-field-driven dynamics~\cite{daquino_frequency-domain_2023}, enabling fast evaluation of magnetization frequency response and absorption spectra from a reduced set of eigenmodes, and has been applied to bridge the gap between macrospin models and full spatial discretization~\cite{perna_computational_2022}.

This document describes the eigenmode solver implemented in the open-source micromagnetic framework \textit{magnum.np}~\cite{bruckner_magnumnp_2023}. Section~\ref{sec:eigenmode} derives the generalized eigenvalue problem, the perturbation treatment of damping, the harmonic excitation response, and the resulting expressions for the magnetization power spectrum and absorbed power. All formulas are expressed using a volume-averaged inner product so that the continuum equations carry over directly to the finite-difference discretization without additional prefactors. Section~\ref{sec:experiments} validates the implementation on two test cases: an elliptical Permalloy nanodisc and an antiferromagnetically coupled bilayer hosting a N\'eel skyrmion.

\subsection{Eigenmode Equation} \label{sec:eigenmode}
The following calculations are based on the work of d'Aquino et al.~\cite{daquino_novel_2009}. Linearization of the Landau-Lifshitz-Gilbert equation(LLG) for small fluctuations $\vec{v}(\vec{x}, t)$ around a stable equilibrium $\vec{m}_0$ leads to
\begin{align} \label{eqn:linearized_LLG}
-\alpha \, \dot{\vec{v}} - \vec{m}_0 \times \dot{\vec{v}} &= \underbrace{\left( \mathbb{1} - \vec{m}_0 \otimes \vec{m}_0 \right)}_{\mathcal{P}_0} \, \underbrace{\gamma \, \left( h_0 \, \mathbb{1} - \vec{h}^\text{lin} \right)}_{\mathcal{A}_0}[\vec{v}],
\end{align}
with the gyromagnetic ratio $\gamma=2.2127615 \cdot 10^5 \si{\metre / \ampere\second}$, the dimensionless damping parameter $\alpha$, the tangent-plane projection operator $\mathcal{P}_0$, the linearized effective field operator $\vec{h}^\text{lin}[\vec{v}] = \frac{\delta \vec{h}^\text{eff}}{\delta \vec{m}}[\vec{v}]$, and parallel component of the equilibrium field $h_0 = \vec{m}_0 \cdot \vec{h}^\text{eff}[\vec{m}_0]$.

Using the Ansatz $\vec{v}(\vec{x}, t) = \sum_k a_k \, \tilde{\vec{\varphi}}'_k(\vec{x}) \, e^{i \omega'_k t}$ with the damped eigenvectors $\tilde{\vec{\varphi}}'_k$ and eigenfrequencies $\omega'_k$, allows to transform the differential equation into an algebraic equation. Due to the linear independence of the basis functions $e^{i \omega'_k t}$ each components $(\omega'_k, \tilde{\vec{\varphi}}'_k)$ needs to fulfill the following algebraic generalized eigenvalue equation:
\begin{align} \label{eqn:eigenvalue_problem_continuum}
\omega'_k \, \underbrace{(-i \alpha \, \tilde{\vec{\varphi}}'_k)}_{i \, \delta \mathcal{B} [\tilde{\vec{\varphi}}'_k]} + \omega'_k \, \underbrace{(-i \vec{m}_0 \times \tilde{\vec{\varphi}}'_k)}_{\mathcal{B}_0 [\tilde{\vec{\varphi}}'_k]} = \underbrace{\mathcal{P}_0 \, \mathcal{A}_0 [\tilde{\vec{\varphi}}'_k]}_{\mathcal{A}_{0\perp} [\tilde{\vec{\varphi}}'_k]}
\end{align}

\subsubsection{Without Damping $\alpha = 0$}
Setting $\alpha = 0$ in Eqn.~\eqref{eqn:eigenvalue_problem_continuum} leads to the following Hermitian generalized eigenvalue problem
\begin{align} \label{eqn:eigenvalue_problem_unperturbated}
\mathcal{A}_{0\perp} \, [\tilde{\vec{\varphi}}_k] = \omega_k \, \mathcal{B}_0 \, [\tilde{\vec{\varphi}}_k],
\end{align}
where $\tilde{\vec{\varphi}}_k$ and $\omega_k$ represent the undamped eigenmodes and eigenfrequencies, respectively. Both operators $\mathcal{A}_{0\perp}$ and $\mathcal{B}_0$ are Hermitian, which allows using the Lanczos algorithm for the solution of the eigenvalue problem, resulting in more stable and efficient calculations. Furthermore it leads to the following orthogonality relation for the undamped eigenmodes (see \cite{daquino_novel_2009} for a detailed prove) which will simplify the calculation of the PSD.

We define the volume-averaged inner product
\begin{equation}\label{eqn:inner_product}
    \langle f, g \rangle = \frac{1}{V}\int f^*\!\cdot g \,\mathrm{d}\vec{x}, \qquad
    \langle f, g \rangle_{\mathcal{O}} = \langle f,\, \mathcal{O}\, g \rangle \qquad
    \| f \|^2 = \langle f, f \rangle
\end{equation}
In the discrete counterpart implemented in {\it magnum.np} is
\begin{equation}\label{eqn:inner_product_discrete}
    \langle f, g \rangle = \frac{1}{N_\mathrm{mag}} \sum_{i \in \Omega_\text{mag}} f_i^* g_i,
\end{equation}
which is implemented as a mean over the magetic domain. Because all continuum formulas are expressed using the volume-averaged inner product $\langle\cdot,\cdot\rangle$, the discrete expressions take exactly the same form, without requiring additional prefactors.

The orthogonality relation finally reads
\begin{align} \label{eqn:orthogonality}
\langle\tilde{\vec{\varphi}}_h, \tilde{\vec{\varphi}}_k\rangle_{\mathcal{A}_{0\perp}} = \omega_h \, \langle\tilde{\vec{\varphi}}_h, \tilde{\vec{\varphi}}_k\rangle_{\mathcal{B}_0} = \delta_{hk}.
\end{align}

\subsubsection{Perturbation analysis for damping $\alpha > 0$}
Starting from Eqn.~\eqref{eqn:eigenvalue_problem_continuum} allowing for small damping $\alpha > 0$ and assuming perturbed quantities $\tilde{\vec{\varphi}}'_k = \tilde{\vec{\varphi}}_k + \delta\tilde{\vec{\varphi}}_k$, $\omega'_k = \omega_k + i \, \delta \omega_k$, and $\mathcal{B}'_0 = \mathcal{B}_0 + i \, \delta \mathcal{B}$ results in the following perturbated eigenvalue problem
\begin{align}
\mathcal{A}_{0\perp} \, \left[ \tilde{\vec{\varphi}}_k + \delta \tilde{\vec{\varphi}}_k \right] = \left( \omega_k + i \, \delta \omega_k \right) \, \left( \mathcal{B}_0 + i \, \delta \mathcal{B} \right) \, \left[ \tilde{\vec{\varphi}}_k + \delta \tilde{\vec{\varphi}}_k \right],
\end{align}
with the perturbed operator $\delta \mathcal{B} = - \alpha \mathbb{1}$. Direct solution of the damped eigenvalue problem can be achieved by means of the Arnoldi method, which also works for non-hermitian eigenvalue problems. However this would influence the solver performance and also destroy orthogonality of the eigenmodes.

Alternatively the perturbation of the eigenvector can be represented using the unperturbed eigenvectors $\delta \tilde{\vec{\varphi}}_h = \sum_k c_{hk} \tilde{\vec{\varphi}}_k$ and a perturbation analysis can be performed. Considering only terms up to first-order perturbations yields
\begin{align} \label{eqn:eigenvalue_problem_alpha}
\mathcal{A}_{0\perp} \, [\delta \tilde{\vec{\varphi}}_k] = \omega_k \, \mathcal{B}_0 \, [\delta \tilde{\vec{\varphi}}_k] + i \, \omega_k \, \delta \mathcal{B} \, [\tilde{\vec{\varphi}}_k] + i \, \delta \omega_k \, \mathcal{B}_0 \, [\tilde{\vec{\varphi}}_k]
\end{align}
Scalar multiplying both sides of equation \eqref{eqn:eigenvalue_problem_alpha} with $\tilde{\vec{\varphi}}_k$ and utilizing the orthogonality relation \eqref{eqn:orthogonality} allows to express the pertubation of the eigenfrequency as
\begin{align}
\delta \omega_k = -\omega_k \frac{\langle\tilde{\vec{\varphi}}_k, \tilde{\vec{\varphi}}_k\rangle_{\delta \mathcal{B}}}{\langle\tilde{\vec{\varphi}}_k, \tilde{\vec{\varphi}}_k\rangle_{\mathcal{B}_0}} = \omega_k^2 \, \langle\tilde{\vec{\varphi}}_k, \alpha \, \tilde{\vec{\varphi}}_k\rangle
\end{align}
Finally the perturbation analysis for small damping results in an additional imaginary contribution to the eigenfrequency, which leads to a damped harmonic oscillation within the time domain. Ignoring the small perturbations of the eigenvectors $\tilde{\vec{\varphi}}_k$ one ends up with
\begin{align}
\vec{v}_\alpha(\vec{x}, t) &\approx \sum_k a_k \, \tilde{\vec{\varphi}}_k(\vec{x}) \, e^{i \omega_k t} \, e^{- \delta \omega_k t}
\end{align}

\subsubsection{Harmonic Excitation}
Adding $\tilde{\vec{h}}^\text{a}$ as a source term to the unperturbed Eqn.~\eqref{eqn:eigenvalue_problem_unperturbated} yields
\begin{align} \label{eqn:eigenvalue_problem_excited}
\omega \, \mathcal{B}_0 \, \delta\tilde{\vec{m}} = \mathcal{A}_{0\perp} \, \delta\tilde{\vec{m}}  - \gamma \, \mathcal{P}_0 \, \tilde{\vec{h}}^\text{a}
\end{align}

In order to utilise the orthogonality relation \eqref{eqn:orthogonality}, the excited solution $\delta\tilde{\vec{m}}$ is expressed by the unperturbed eigenvectors
\begin{align} \label{eqn:v_th_decomposition}
\delta\tilde{\vec{m}}(\vec{x}, \omega) = \sum_k \tilde{a}_k(\omega) \, \tilde{\vec{\varphi}}_k(\vec{x})
\end{align}

Scalar multiplying both sides of equation \eqref{eqn:eigenvalue_problem_excited} with $\tilde{\vec{\varphi}}_h$ and utilizing the orthogonality relation allows to calculate the mode amplitudes $\tilde{a}_k(\omega)$
\begin{align}
\omega \, \underbrace{\left\langle\tilde{\vec{\varphi}}_h, \sum \tilde{a}_k \, \tilde{\vec{\varphi}}_k\right\rangle_{\mathcal{B}_0}}_{\frac{1}{\omega_h} \tilde{a}_h} &= \underbrace{\left\langle\tilde{\vec{\varphi}}_h, \sum \tilde{a}_k \, \tilde{\vec{\varphi}}_k\right\rangle_{\mathcal{A}_{0\perp}}}_{\tilde{a}_h} - \, \gamma \, \underbrace{\left\langle\tilde{\vec{\varphi}}_h, \mathcal{P}_0 \, \tilde{\vec{h}}^\text{a} \right\rangle}_{\tilde{h}_h},
\end{align}
which finally results in the following expression for the mode amplitudes
\begin{align}\label{eqn:mode_expansion}
\tilde{a}_h = \frac{\gamma \, \omega_h}{\omega_h' - \omega} \, \tilde{h}_h
\quad \Rightarrow \quad
\delta\tilde{\vec{m}}(\vec{x}, \omega) = \sum_k \frac{\gamma \, \omega_k}{\omega_k' - \omega} \, \tilde{h}_k \, \tilde{\vec{\varphi}}_k(\vec{x})
\end{align}
where the previously calculated damped eigenfrequencies $\omega_h'$ are used in the denominator.

\subsubsection{Power Spectrum of Magnetization}
The volume-averaged power spectrum of the magnetization driven by a harmonic excitation $\tilde{\vec{h}}^\text{a}$ is defined as (compare Eqn.~(22) of Ref.~\cite{daquino_frequency-domain_2023})
\begin{align}\label{eqn:spectrum}
    \tilde{p}(\omega) &= \frac{1}{2V} \int M_s^2 \left|\delta\tilde{\vec{m}}(\vec{x}, \omega)\right|^2 \dx \\
                      &= \frac{1}{2} \| M_s \delta\tilde{\vec{m}}(\vec{x}, \omega)\|^2,
\end{align}
with $\delta\tilde{\vec{m}}$ being the harmonic magnetization response driven by $\tilde{\vec{h}}^\text{a}$.
Inserting the modal expansion \eqref{eqn:mode_expansion} yields
\begin{align}
    \tilde{p}(\omega) &= \frac{\gamma^2}{2} \sum_{k,h} \frac{\omega_k \omega_h}{(\omega_k' - \omega)(\omega_h'^* - \omega)}\, \tilde{h}_k \tilde{h}_h^* \langle M_s \tilde{\vec{\varphi}}_k, M_s \tilde{\vec{\varphi}}_h \rangle \\
    &\approx \frac{\gamma^2}{2} \sum_k \frac{\omega_k^2 \, |\tilde{h}_k|^2 \, \| M_s \tilde{\vec{\varphi}}_k\|^2}{(\omega_k - \omega)^2 + \delta \omega_k^2} \\
    &= \frac{1}{2} \sum_k |\tilde{a}_k|^2 \, \| M_s \tilde{\vec{\varphi}}_k\|^2
\end{align}
where the off-diagonal terms ($k \neq h$) have been neglected, which is justified for well separated resonances.

\subsubsection{Absorbed Magnetic Power}
The average power absorbed by the magnetic system $P_\text{abs}(\omega)$ driven by a harmonic excitation $\tilde{\vec{h}}^\text{a}$ can be defined as the real part of the complex absorbed power $\hat{P}_\text{abs}(\omega)$ (compare Eq.~(25) of Ref.~\cite{daquino_frequency-domain_2023})
\begin{align}
  \hat{P}_\text{abs}(\omega) &= \frac{\mu_0}{2V} \int i \omega \, M_s(\vec{x}) \,\delta\tilde{\vec{m}}(\vec{x}, \omega) \cdot \tilde{\vec{h}}^{\text{a}*}(\vec{x}) \,\mathrm{d}\vec{x} \\
                             &= \frac{\mu_0}{2}\, i\omega \, \langle \tilde{\vec{h}}^\text{a},\, M_s \,\delta\tilde{\vec{m}} \rangle.
\end{align}
Inserting the modal expansion \eqref{eqn:mode_expansion} and defining the $M_s$-weighted modal projection $\tilde{g}_k = \langle\tilde{\vec{\varphi}}_k, M_s \mathcal{P}_0 \, \tilde{\vec{h}}^\text{a}\rangle$ yields
\begin{align}
    \hat{P}_\text{abs}(\omega) &= \frac{\mu_0}{2} \sum_k i \omega \, \tilde{a}_k \, \tilde{g}_k^{*} \label{eqn:Pabs_modal}\\
    &= \frac{\mu_0 \, \gamma}{2} \sum_k \frac{i \omega \, \omega_k \, \tilde{h}_k \, \tilde{g}_k^{*}}{\omega_k - \omega + \delta \omega_k}.
\end{align}

\subsection{Numerical Experiments} \label{sec:experiments}

\subsubsection{Elliptical nanodisc (\texttt{run\_carlotti.py})}
The eigenmode method is first applied to an elliptical Permalloy nanodisc with dimensions $\SI{200}{nm} \times \SI{100}{nm} \times \SI{5}{nm}$, discretized on a $200 \times 100 \times 1$ finite-difference grid with a cell size of $\SI{1}{nm} \times \SI{1}{nm} \times \SI{5}{nm}$. The elliptical shape is defined by masking cells outside the ellipse boundary. The material parameters are $M_s = \SI{800}{kA/m}$, $A = \SI{13}{pJ/m}$, and $\alpha = 0.01$. The effective field includes exchange and demagnetization contributions.

\begin{figure}
  \centering
  \includegraphics[width=\columnwidth]{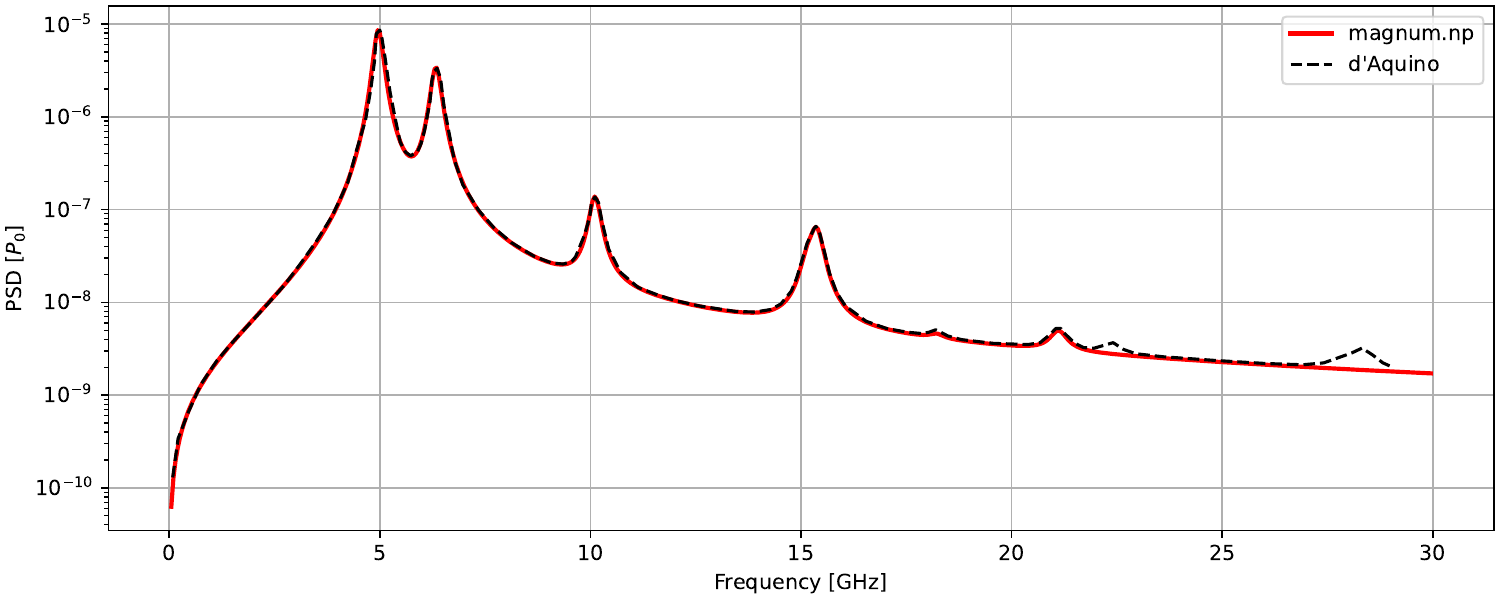}
  \caption{Absorbed magnetic power of the elliptical Permalloy nanodisc computed with the eigenmode method (\texttt{run\_carlotti.py}, solid red) compared with reference data from d'Aquino et al.\ (dashed black).}
  \label{fig:result_carlotti}
\end{figure}

The equilibrium magnetization is obtained by energy minimization using the Barzilai--Borwein method. The first 20 eigenmodes are then computed by solving the generalized eigenvalue problem~\eqref{eqn:eigenvalue_problem_unperturbated} with the Lanczos algorithm. The absorbed power $P_\text{abs}(\omega)$ is evaluated semi-analytically from the modal expansion using a uniform in-plane excitation field $\tilde{\vec{h}}^\text{a} = (0,\, \SI{0.5}{mT}/\mu_0,\, 0)$. The eigenmode-based result is compared with reference data from d'Aquino et al.~\cite{daquino_frequency-domain_2023} in Fig.~\ref{fig:result_carlotti}.

\subsubsection{Antiferromagnetically coupled bilayer (\texttt{run\_bilayer.py})}
The second example considers an antiferromagnetically coupled bilayer with dimensions $\SI{120}{nm} \times \SI{120}{nm} \times \SI{10}{nm}$, discretized on a $24 \times 24 \times 2$ grid (cell size $\SI{5}{nm} \times \SI{5}{nm} \times \SI{5}{nm}$). Each layer occupies one cell along the $z$-direction. The material parameters are $M_s = \SI{800}{kA/m}$, $A = \SI{13}{pJ/m}$, interfacial DMI $D_i = \SI{-3}{mJ/m^2}$, uniaxial anisotropy $K_u = \SI{0.5}{MJ/m^3}$ along $z$, RKKY interlayer coupling with $J_\text{RKKY} = \SI{-0.3}{mJ/m^2}$, and $\alpha = 0.008$. A static bias field and a N\'eel skyrmion texture are used to initialize the magnetization, with opposite core polarities in the two layers. The equilibrium state is obtained by energy minimization.

\begin{figure}
  \centering
  \includegraphics[width=\columnwidth]{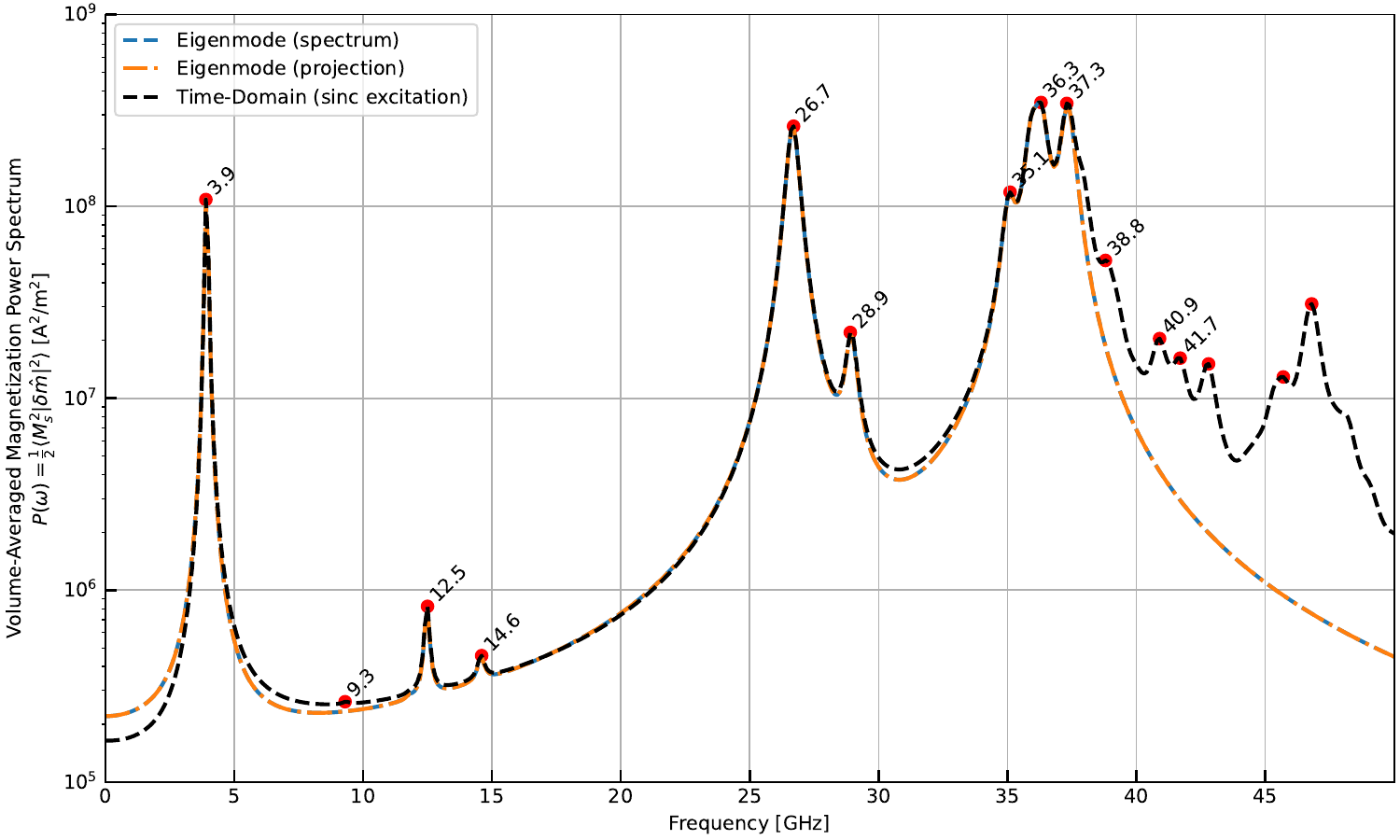}
  \caption{Volume-averaged magnetization power spectrum of the antiferromagnetically coupled bilayer hosting a N\'eel skyrmion. The eigenmode-based spectrum (dashed) and projection (dash-dotted) from \texttt{run\_bilayer.py} are compared with the time-domain sinc-excitation result (black dashed).}
  \label{fig:result_bilayer}
\end{figure}

The first 20 eigenmodes are computed from the generalized eigenvalue problem. Both the eigenmode-based power spectrum $\tilde{p}(\omega)$ and the modal projection are compared with a time-domain simulation using a sinc-pulse excitation ($f_\text{max} = \SI{100}{GHz}$, total simulation time $\SI{10}{ns}$, time step $\Delta t = \SI{2}{ps}$). Results are shown in Fig.~\ref{fig:result_bilayer}.

\subsection{Conclusion} \label{sec:conclusion}
We have presented the eigenmode solver of \textit{magnum.np}, which computes spin-wave resonances and their spectral signatures directly from a linearization of the LLG equation. By formulating all continuum equations with a volume-averaged inner product, the analytical expressions for the power spectrum and absorbed power translate one-to-one to the finite-difference implementation without rescaling. The method was validated on an elliptical Permalloy nanodisc, where the absorbed power agrees with published reference data, and on an antiferromagnetically coupled bilayer with a N\'eel skyrmion, where the eigenmode-based spectrum reproduces the time-domain simulation result. In both cases the eigenmode approach provides smooth, noise-free spectra at a fraction of the computational cost of stochastic time-integration.

\bibliography{supplementary-eigenmode-solver}